\documentclass[conference]{IEEEtran}
\IEEEoverridecommandlockouts
\usepackage{cite}
\usepackage{algorithmic}
\usepackage{textcomp}
\usepackage{xcolor}
\def\BibTeX{{\rm B\kern-.05em{\sc i\kern-.025em b}\kern-.08em
    T\kern-.1667em\lower.7ex\hbox{E}\kern-.125emX}}
\usepackage{multicol}
\usepackage{graphicx}
\usepackage{subfigure}
\usepackage{caption}
\usepackage{pifont}
\usepackage{amsmath,amsfonts,amssymb}
\usepackage{mathtools}
\usepackage{enumitem}
\usepackage{multirow}

\usepackage{titlesec}
\titlespacing\section{2pt}{2pt plus 2pt minus 2pt}{0pt plus 1pt minus 1pt}
\titlespacing\subsection{2pt}{1pt plus 2pt minus 1pt}{0pt plus 1pt minus 1pt}
\titlespacing\subsubsection{2pt}{1pt plus 2pt minus 1pt}{0pt plus 1pt minus 1pt}

\begin{document}

\title{\LARGE \bf FindeR: Accelerating FM-Index-based Exact Pattern Matching in Genomic Sequences through ReRAM technology}

\author{\IEEEauthorblockN{Farzaneh Zokaee$\ddag$}
\IEEEauthorblockA{\textit{Indiana University Bloomington} \\
\texttt{fzokaee@iu.edu}}
\and
\IEEEauthorblockN{Mingzhe Zhang}
\IEEEauthorblockA{\textit{ICT, CAS, China} \\
\texttt{zhangmingzhe@ict.ac.cn}}
\and
\IEEEauthorblockN{Lei Jiang$\ddag$\thanks{$\ddag$This work was supported in part by NSF CCF-1909509.}}
\IEEEauthorblockA{\textit{Indiana University Bloomington} \\
\texttt{jiang60@iu.edu}}
}

\maketitle

\begin{abstract}
Genomics is the critical key to enabling precision medicine, ensuring global food security and enforcing wildlife conservation. The massive genomic data produced by various genome sequencing technologies presents a significant challenge for genome analysis. Because of errors from sequencing machines and genetic variations, approximate pattern matching (APM) is a must for practical genome analysis. Recent work proposes FPGA, ASIC and even process-in-memory-based accelerators to boost the APM throughput by accelerating dynamic-programming-based algorithms (e.g., Smith-Waterman). However, existing accelerators lack the efficient hardware acceleration for the exact pattern matching (EPM) that is an even more critical and essential function widely used in almost every step of genome analysis including assembly, alignment, annotation and compression. 

State-of-the-art genome analysis adopts the FM-Index that augments the space-efficient BWT with additional data structures permitting fast EPM operations. But the FM-Index is notorious for poor spatial locality and massive random memory accesses. In this paper, we propose a ReRAM-based process-in-memory architecture, FindeR, to enhance the FM-Index EPM search throughput in genomic sequences. We build a reliable and energy-efficient Hamming distance unit to accelerate the computing kernel of FM-Index search using commodity ReRAM chips without introducing extra CMOS logic. We further architect a full-fledged FM-Index search pipeline and improve its search throughput by lightweight scheduling on the NVDIMM. We also create a system library for programmers to invoke FindeR to perform EPMs in genome analysis. Compared to state-of-the-art accelerators, FindeR improves the FM-Index search throughput by $83\%\sim 30K\times$ and throughput per Watt by $3.5\times\sim 42.5K\times$.
\end{abstract}

\begin{IEEEkeywords}
short DNA alignment, ReRAM
\end{IEEEkeywords}

\section{Introduction}
\label{sec:introduction}

High throughput sequencing technologies (i.e., Illumina~\cite{Schirmer:BMCB2016}, PacBio SMRT~\cite{Mosher:JMM2014} and Oxford Nanopore~\cite{Eisenstein:Oxford2012}) have revolutionized biological sciences, since they can sequence an entire human genome within a single day. The explosion of genomic data has been the cornerstone in enabling the understanding of complex human diseases~\cite{Merker:Nature2018}, ensuring global food security~\cite{Ma:TBIO2017} and enforcing wildlife conservation~\cite{Arif:SJBS2011}. Several government projects~\cite{Hanna:JMAHP2017} around the world have been launched to deploy whole genome sequencing in clinical practice and public health. Genome analysis will become one of the standard practices of newborn screening over the next decade. The \textit{speed} of genome analysis of big genomic data is a matter of life and death.

However, it is \textit{challenging} to efficiently store, process and analyze the huge amounts of genomic data generated by sequencing machines that translate organic nucleotides to digital symbols. The genome sequencing machines are improving faster than Moore's Law~\cite{Canzar:IEEE2017}. For instance, a recent Illumina NovaSeq machine~\cite{Davies:COMICS2017} produces nearly 750GB of data per day. Oxford Nanopore even creates USB-drive-style devices to sequence organisms in the wild~\cite{Eisenstein:Oxford2012}. It is projected that by 2025 the total amount of genomic data on earth will exceed the data capacity of YouTube and Twitter~\cite{Zachary:PLOS2015}. As a result, the exponential genomic data growth significantly increases pressure on hardware platforms for genome analysis. Analyzing a single genome may take hundreds of CPU hours~\cite{Turakhia:ASPLOS2018,Fuijiki:ISCA2018} on high-end servers. Application-specific acceleration for genome analysis has become essential. 

\begin{figure*}[htbp!]
\centering
\includegraphics[width=\linewidth]{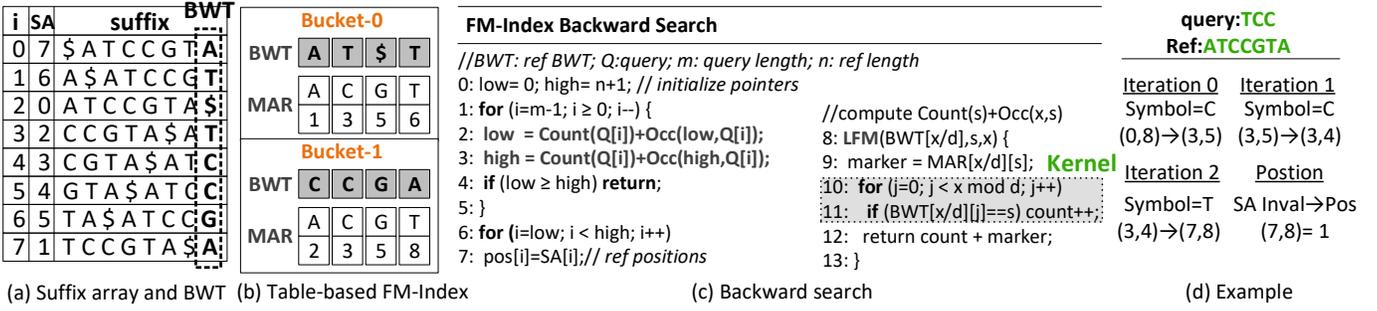}
\caption{The FM-Index overview: (a) SA and BWT; (b) FM-Index; (c) Backward search; (d) An example.}
\label{f:fm_index_all}
\vspace{-0.2in}
\end{figure*}

Critical steps in genome analysis such as \textit{assembly} and \textit{alignment} involve both exact and approximate pattern matching~\cite{Pevsner:BFG2015,Canzar:IEEE2017}, because of their \textit{seed-and-extend} paradigm~\cite{Wang:BMC2018,Huang:BMC2017,Simpson:GR2012,Li:BWAMEM2013,Li:BIOINFO2012}. Parts of a genome fragment (called seeds) are mapped to their matched positions in other fragments (assembly) or long references (alignment) by \textit{exact pattern matching} (EPM) during seeding. Seed extension pieces together a larger sequence with seed mappings and edit distance errors, i.e., insertions, deletions (indels) and substitutions, by \textit{approximate pattern matching} (APM). All genome fragments have to go through the seeding stage, but only the fragments containing at least one exactly matched seed are actually processed in the seed extension phase. So EPM operations during seeding cost even larger amounts of CPU time than APM operations during seed extension in state-of-the-art genome analysis software~\cite{Houtgast:SAMOS2015,Chang:FCCM2016}. Recent accelerators heavily optimize APM in seed extension by accelerating dynamic programming-based algorithms~\cite{Turakhia:ASPLOS2018} (e.g., Smith-Waterman algorithm~\cite{Kaplan:MICRO2017,Madhavan:ISCA2014,Enzo:ICBB2017}) or Universal Levenshtein Automata~\cite{Fuijiki:ISCA2018}, because APM handles edit distance errors and thus has higher time complexity. In contrast, for exact pattern matching (EPM) during seeding, these accelerators~\cite{Kaplan:MICRO2017,Karen:CORR2017,Madhavan:ISCA2014,Turakhia:ASPLOS2018,Fuijiki:ISCA2018,Enzo:ICBB2017,Huangfu:DAC2018} use a simple hash table to find positions exactly matching a seed (hit) in a sequence. However, the hash-table-based technique is inefficient to implement EPM during seeding. For high sensitivity and precision, a huge number of seeds have to be used to produce large amounts of candidate positions with seed hits. The seeds and candidate positions may occupy $80GB$~\cite{Turakhia:ASPLOS2018} DRAM. Furthermore, false positive candidate positions generated by the hash-table-based technique slow down the seed extension.

The Ferragina-Manzini Index (FM-Index)~\cite{Ferragina:SODA2001} is adopted by state-of-the-art genome analysis software such as BWA-MEM~\cite{Li:BWAMEM2013} and SOAP~\cite{Luo:PLOS2013} to build super-maximal exact matches (SMEMs) during seeding, since it augments the space-efficient Burrows-Wheeler transform~\cite{Burrows:HSRR1994} with accessory data structures that permit fast EPM. SMEMs generated by the FM-Index guarantee each seed does not overlap other seeds and has the maximal length that cannot be further extended. Compared to the hash-table-based technique, the FM-Index reduces not only the number of errors in final genome mappings but also the seed extension duration~\cite{Ahmed:ICBB2016}. 

Besides genome assembly and alignment, FM-Index is also widely used in other time-consuming steps of genome analysis such as genome annotation~\cite{Healy:GEN2003} and compression~\cite{Prochazka:DCC2014,Arram:FPT2015} for EPM. Recent work presents FPGA~\cite{Chang:FCCM2016,Houtgast:SAMOS2015,Arram:FPT2015,Arram:TCBB2017} and ASIC~\cite{Wu:ITBCS2017}-based accelerators to process FM-Index searches. However, the FM-Index is notorious for its massive random memory accesses~\cite{Chacon:TCBB2015,Chacon:PCS2013}, so it hits the \textit{memory wall} of the \textit{von Neumann} architecture. Existing FM-Index accelerators are fundamentally limited by memory bandwidth. In this paper, we propose a ReRAM-based process-in-memory architecture, FindeR, to accelerate FM-Index-based exact pattern matching (EPM) in genomic sequences. Our contributions are summarized as follows.
\begin{itemize}[nosep,leftmargin=*]
\item \textbf{A ReRAM-based Hamming distance unit --} We transform a symbol-counting operation in the FM-Index search kernel to a Hamming distance calculation. We then propose a ReRAM-based Hamming distance unit (RHU) and a lookup-table-based adder to accelerate FM-Index searches in commodity ReRAM chips without extra CMOS gates.

\item \textbf{Reliability, power and density --} We present architectural techniques to provide RHUs long lifetime and high current accumulation accuracy. To balance the trade-off between area, power and energy, we comprehensively explore the design space of FM-Index by tuning its bucket size.

\item \textbf{A full-fledged pipeline with system support --} We further architect a full-fledged FM-Index pipeline and improve the search throughput by lightweight scheduling. Finally, we design a system-level interface for genomics developers to run FindeR for EPM. 

\item \textbf{EPM throughput per Watt --} We evaluate and compare FindeR to prior hardware platforms in various genome analysis applications such as genome assembly, alignment, annotation and compression. The results show that our PIM improves the FM-Index search throughput by $83\%\sim 30K\times$ and throughput per Watt by $3.5\times\sim 42.5K\times$ over the state-of-the-art accelerators.
\end{itemize}

\section{Background and Motivation}
\label{s:back}

\subsection{Exact Pattern Matching in Genome Analysis}

\subsubsection{Genome Analysis}

Genome analysis~\cite{Pevsner:BFG2015} mainly includes four steps: \textit{sequencing}, \textit{assembly}, \textit{alignment} and \textit{annotation}. In genome sequencing, sequencing machines translate organic nucleotide fragments to digital DNA sequences (called \textit{read}s) comprising $A$, $C$, $G$ and $T$. Assembly constructs larger DNA sequences by merging different reads, while alignment decides the precise order of nucleotides by aligning short reads to a long reference genome. Finally, annotation attaches biological information to long sequences.

\subsubsection{Exact Pattern Matching}

Genomic data have two significant sources of errors: sequencing machine errors and true variations~\cite{Quail:BMC2012}. The sequencing error rate of various sequencing machines is $0.2\%\sim30\%$~\cite{Quail:BMC2012}, while the overall variation of the human population has been estimated as $0.1\%$~\cite{Canzar:IEEE2017}. So practical genome analysis has to support approximate pattern matching (APM) to accommodate errors. The APMs are often implemented by the Smith-Waterman (SW) algorithm~\cite{Kaplan:MICRO2017,Madhavan:ISCA2014,Enzo:ICBB2017} that has high time complexity and is impractically slow. To avoid the prohibitive SW overhead, the seed-and-extension paradigm is adopted in genome assembly~\cite{Li:BIOINFO2012,Turakhia:ASPLOS2018,Simpson:GR2012} and alignment~\cite{Canzar:IEEE2017,Li:BWAMEM2013}. The seeding stage heavily depends on Exact Pattern Matching (EPM) to find parts of a read that exactly match at least one position in another read or the reference. Compared to an APM, an EPM is more elementary, since an APM can be broken into multiple EPMs~\cite{Canzar:IEEE2017}. The FM-Index~\cite{Healy:GEN2003,Prochazka:DCC2014,Arram:TCBB2017,Wang:ICPP2018} is one of the most computationally and memory efficient solutions to performing EPM.


\subsection{FM-Index}
\label{s:fm_index_all}

\subsubsection{Data Structure}
The Ferragina-Manzini Index (FM-Index) \cite{Ferragina:SODA2001} is built upon the Burrows-Wheeler transform (BWT) \cite{Burrows:HSRR1994}, a permutation of a symbol sequence generated from its suffix array (SA). The SA of a reference genome $\mathcal{G}$ is a lexicographically sorted array of the suffixes of $\mathcal{G}$, where each suffix is denoted by its position in $\mathcal{G}$. $\$$ indicates the end of a genome, so it is in the lexicographically smallest position. Each position of the BWT is calculated by 
\vspace{-0.1in}
\begin{equation}
\centering
BWT[i]=\mathcal{G}[(SA[i]-1) \bmod |\mathcal{G}|],
\label{e:dna_fm_bwt}
\end{equation}
where $|\mathcal{G}|$ is the length of $\mathcal{G}$. Figure~\ref{f:fm_index_all}(a) shows the SA of a reference sequence $\mathcal{G}=ATCCGTA\$$ and its $BWT(\mathcal{G})=AT\$TCCGA$. The FM-Index implements EPM searches through two functions $Count(s)$ and $Occ(s,i)$ performed on the BWT. $Count(s)$ computes the number of symbols in the BWT that are lexicographically smaller than the symbol $s$, e.g., $Count(T)=6$. $Occ(s,i)$ returns the number of symbol $s$ in the BWT from positions 0 to $i-1$, e.g., $Occ(C,5)=1$. The values of $Count(s)$ and $Occ(s,i)$ can be pre-calculated and stored, but the storage overhead is significant. To keep the FM-Index size in check, the $Occ$ values are sampled into buckets of width $d$ shown in Figure~\ref{f:fm_index_all}(b) ($d=4$). The $Occ$ values are stored each $d$ positions as markers (MARs) to reduce the storage overhead by a factor of $d$. The omitted $Occ$ values can be reconstructed by summing the previous marker and the number of symbol $s$ from the remaining positions in the BWT bucket. We call this data structure {\it table}-based FM-Index. To simplify search operations, the $Count$ values for each symbol are added to the corresponding markers. The markers and the BWT buckets are interleaved to build the FM-Index. The FM-Index size ($F$) can be calculated by Equation~\ref{e:dna_fm_size}, where $|\Sigma|$ is the alphabet size. The first item in the equation indicates the $Occ$ table size, while the second represents the BWT size. For a human genome ($|\mathcal{G}|=3G$, $|\Sigma|=4$ and $d=128$), the FM-Index costs 1.5GB.
\begin{equation}
\centering
F=\frac{4\cdot |\mathcal{G}|\cdot |\Sigma|}{d} + \frac{|\mathcal{G}| \cdot \lceil log_2(|\Sigma|+1) \rceil}{8}\quad bytes
\label{e:dna_fm_size}
\end{equation}

\begin{figure}[t!]
\centering
\includegraphics[width=\linewidth]{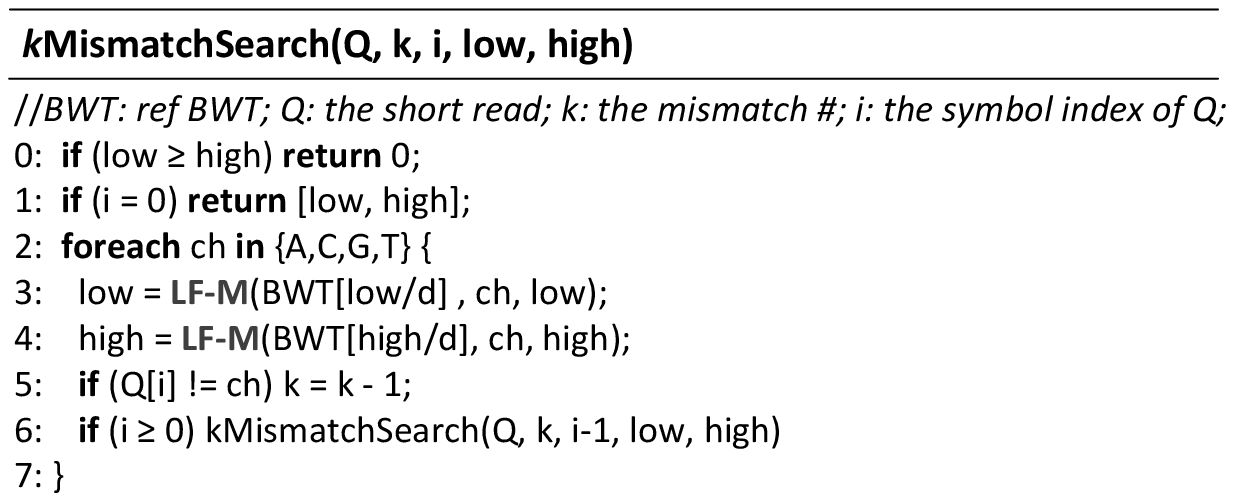}
\vspace{-0.2in}
\caption{The FM-Index-based $k$-mismatch search.}
\label{f:dna_fm_kmismatch}
\vspace{-0.3in}
\end{figure}

\subsubsection{Backward Search}

EPM is achieved by the FM-Index backward search, whose algorithm can be viewed in Figure~\ref{f:fm_index_all}(c). The SA interval $(low, high)$ covers a range of indices in the SA where the suffixes have the same prefix. The pointer $low$ locates the index in the SA where the pattern is first found as a prefix, while the pointer $high$ provides the index after the one where the pattern is last found. At first, $low$ and $high$ are initialized to the minimum and maximum indices of the $Occ$ array respectively. And then, iterating from the last symbol in the read to the first, the SA interval is updated using the Last-First Mapping (\textit{LFM}) function where $low$ is calculated:
\begin{equation}
\centering
Count(Q[i])+Occ(Q[i],low).
\label{e:dna_fm_lpm}
\end{equation}
$high$ can be computed in the same way. The function of \textit{LFM} is shown from line 8 to 13 in Figure~\ref{f:fm_index_all}(c), where the computations of $low$ and $high$ are \textbf{random pointer chasing having poor spatial locality}. Finally, the SA interval gives the range of indices in the SA where the suffixes have the target read as a prefix. These indices are converted to reference genome positions using the SA. Figure~\ref{f:fm_index_all}(d) illustrates an example of searching $TCC$ in the reference $\mathcal{G}=ATCCGTA$. Before a search happens, $(low, high)$ is initialized to $(0,8)$. In Iteration 0, the last symbol $C$ is processed by \textit{LFM} where $(low, high)$ is updated to $(3,5)$. After three iterations, $(low, high)$ eventually equals $(7,8)$. By looking up $SA[7]$ in Figure~\ref{f:fm_index_all}(a), we find that the read $TCC$ in reference $\mathcal{G}=ATCCGTA$ starts from $\mathcal{G}[1]$. Once the FM-Index is built, huge volumes of backward searches are invoked by almost every step of genome analysis.

\subsubsection{The FM-Index for $k$-Mismatch Search}

Compared to an APM, an EPM is more elementary, since an APM can be transformed into multiple EPMs to handle indels and substitutions~\cite{Canzar:IEEE2017}. However, the FM-Index-based $k$-error search that is able to handle both indels and substitutions exponentially increases the number of iterations in the \textit{LFM} function. So the seed extension phase of long read alignment whose major error type is indel uses only the SW algorithm~\cite{Turakhia:ASPLOS2018}. On the contrary, the FM-Index-based $k$-mismatch search that can accommodate only substitutions moderately increases the iteration number of the \textit{LFM} function. Considering that 98.9\% of short read sequencing errors~\cite{Quail:BMC2012} are mismatches, the state-of-the-art aligners such as BWA-MEM~\cite{Li:BWAMEM2013} and SOAP~\cite{Luo:PLOS2013} implement the \textbf{APMs} to process mismatches in the seed extension phase of short read alignment by \textbf{$k$-mismatch FM-Index searches}~\cite{Canzar:IEEE2017}. 

The algorithm of $k$-mismatch FM-Index search is shown in Figure~\ref{f:dna_fm_kmismatch}. Its main difference from the normal FM-index backward search (Figure~\ref{f:fm_index_all}(c)) lies in line 2, where in addition to the current symbol of the read, all possible substitutions are tested until the substitution number exceeds $k$. To search in both directions, the FM-Index is generated for both the BWT and its reverse~\cite{Lam:ICBB2009}. The bi-directional FM-Index searches from the middle of a read, uses the forward search for the first half, and adopts the backward search for the second half. The iteration number of $k$-mismatch FM-Index is decided by both the read length and the mismatch positions.

\subsubsection{FM-Index Applications}
\label{s:fm_app}

The FM-Index has been widely adopted for EPM in various steps of genome analysis: 
\begin{itemize}[nosep,leftmargin=*]

\item \textbf{SMEM Seeding}. The state-of-the-art genome aligners, e.g., BWA-MEM~\cite{Li:BWAMEM2013} and SOAP~\cite{Luo:PLOS2013}, create super-maximal exact matches (SMEMs) during seeding by FM-Index. A SMEM is an exact match in a read that cannot be extended further in either direction or contained in any other SMEM. Compared to hash-table-based seeding, SMEM seeding reduces not only the number of errors in genome mappings but also seed extension duration~\cite{Ahmed:ICBB2016}. The detailed algorithm for SMEM construction can be found in~\cite{Li:BIOINFO2012}. The construction of SMEM first extends the seed rightward by FM-Index searches. The SA intervals of the hits sharing the same starting position are stored in an interval set. The construction of SMEM further leftward extends the hits found in rightward seed extensions by FM-Index searches too. The FM-Index search has been identified as the most time-consuming operation for SMEM constructions~\cite{Ahmed:ICCAD2015}. 

\item \textbf{Seed Extension for Short Reads}. The error rate of Illumina sequencing machines for short ($100$-bp) reads is only $\sim0.2\%$~\cite{Quail:BMC2012}. Rather than indels, the majority of the Illumina errors are substitutions~\cite{Quail:BMC2012} (mismatches). The overall variation in the human population is $0.1\%$~\cite{Canzar:IEEE2017}. $74\%$ of $100$-bp reads have no mismatch. $22.4\%$ have 1 mismatch, while $3.3\%$ have 2 mismatches. The state-of-the-art aligners~\cite{Li:BWAMEM2013,Luo:PLOS2013} use $k$-mismatch bi-directional FM-Index searches for short read seed extensions. 99.7\% of short reads~\cite{Arram:TCBB2017} are actually aligned by the $2$-mismatch bi-directional FM-Index.

\item \textbf{Genome Annotation}. A keyword comprising multiple symbols in $\Sigma$ has a certain number of exact matches within a genome. Keyword counting is indispensable in annotation applications such as probe design, discovery of repeat elements, and mathematical modeling of genome evolution. The FM-Index search is the key operation for computing the keyword count~\cite{Healy:GEN2003}.

\item \textbf{Genome Compression}. The genomic big data explosion produced by emerging sequencing technologies poses a profound storage challenge. Biological sequences of the same species are highly similar and differ only by single nucleotide polymorphisms. The reference-based compression algorithm~\cite{Prochazka:DCC2014} records only the differences between a genome sequence and the reference, so it achieves higher compression ratios than general purpose compressors including gzip and bzip2~\cite{Arram:FPT2015}. The FM-Index searches deciding the mapping to the reference account for $70\%$ of the compression time~\cite{Prochazka:DCC2014}.
\end{itemize}

\begin{figure}[htbp!]
\vspace{-0.1in}
\centering
\includegraphics[width=0.95\linewidth]{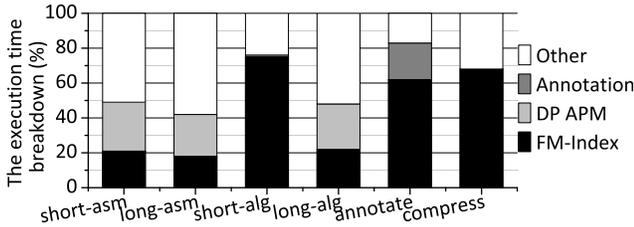}
\vspace{-0.1in}
\caption{Execution time breakdown in genome analysis.}
\label{f:dna_fm_mot}
\vspace{-0.1in}
\end{figure}

Figure~\ref{f:dna_fm_mot} shows the execution time breakdown of genomic applications. Our experimental methodology is elaborated in Section~\ref{s:eandm}. On average, \textbf{FM-Index searches cost $18\%\sim75\%$ of the total execution time in these applications}. Particularly, the FM-Index search is heavily invoked in short read alignment, since SMEM seeding and most seed extensions rely on it. In short read alignment, the FM-Index-based EPM consumes significantly more CPU time than that spent by the dynamic programming (DP) APM. APMs spend $24\%\sim28\%$ of the total execution time in assembling short and long reads, and aligning long reads. But the other steps of genome analysis hardly use FM-Index searches.

\subsection{ReRAM for Big Genomic Data}

\subsubsection{Processing Genomic Data in ReRAM}

Because of its high density, long cell endurance, and low write power consumption, ReRAM~\cite{Karen:CORR2017,Kaplan:MICRO2017,Huangfu:DAC2018} is deemed one of the most promising nonvolatile memory technologies to overcome the big genomic data challenge. Through the $\mathbf{4F^2}$ cell size, multi-level cells and cross-point array structure, a ReRAM-based main memory maintains scalable performance for genome analysis applications. However, recent work uses low-density and power-hungry ReRAM-based content address memories (CAMs) to perform hash-based short read seeding~\cite{Karen:CORR2017,Huangfu:DAC2018} or dynamic-programming-based APMs~\cite{Kaplan:MICRO2017} for short read seed extensions. Compared to APM, EPM consumes even more time in various genomic applications. However, \textbf{no prior work focuses on accelerating FM-Index-based EPM operations by ReRAM technology}. 

\begin{figure}[ht!]
\vspace{-0.1in}
\centering
\subfigure[Cell.]{
   \includegraphics[width=0.45\columnwidth]{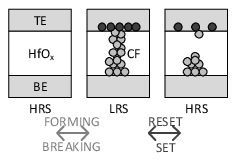}
   \label{f:reram_mem_cell}
}
\hspace{-0.15in}
\subfigure[Switching.]{
   \includegraphics[width=0.4\columnwidth]{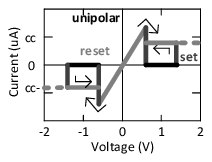}
   \label{f:reram_polar_go}
}
\vspace{-0.1in}
\caption{The unipolar switching ReRAM.}
\label{f:reram_polar_switching}
\vspace{-0.2in}
\end{figure}

\subsubsection{Basics}

A ReRAM cell storing data in a thin layer of metal-oxide ($HfO_{x}$) is sandwiched by a top electrode (TE) and a bottom electrode (BE), as shown in Figure~\ref{f:reram_mem_cell}. By injecting electrical pulses, it can be switched to a high resistance state (HRS) or a low resistance state (LRS). The transition from HRS to LRS is called \textit{SET}, while the reverse is named \textit{RESET}. To initiate the switching between HRS and LRS in a metal-oxide layer, a ReRAM cell needs a conductive filament (CF) created to connect the TE and BE by a high voltage \textit{FORMING} operation. Without the CF, the cell cannot be RESET or SET, but stays only in HRS. A RESET yields a HRS cell by rupturing cracks on the CF, while a SET recovers the complete CF resulting in a LRS cell. Through a \textit{BREAKING} operation ($\sim100\mu s$), the CF in a cell can be eliminated. There are two methods to implement ReRAM writes~\cite{Wong:IEEE2012}: \textit{bipolar switching} and \textit{unipolar switching}. ReRAM cells are connected by bit-lines (BLs) and word-lines (WLs) to form an array. We define the voltage difference generated by a high (low) WL voltage and a low (high) BL voltage as the positive (negative) voltage. Bipolar switching sets cells by positive voltages and resets cells by negative voltages. On the contrary, as Figure~\ref{f:reram_polar_go} exhibits, during unipolar switching, both SETs and RESETs are completed with either positive or negative voltages~\cite{Ning:SR2017}. 

\begin{figure*}[t!]
\centering
\begin{minipage}{.65\columnwidth}
   \centering
   \includegraphics[width=\columnwidth]{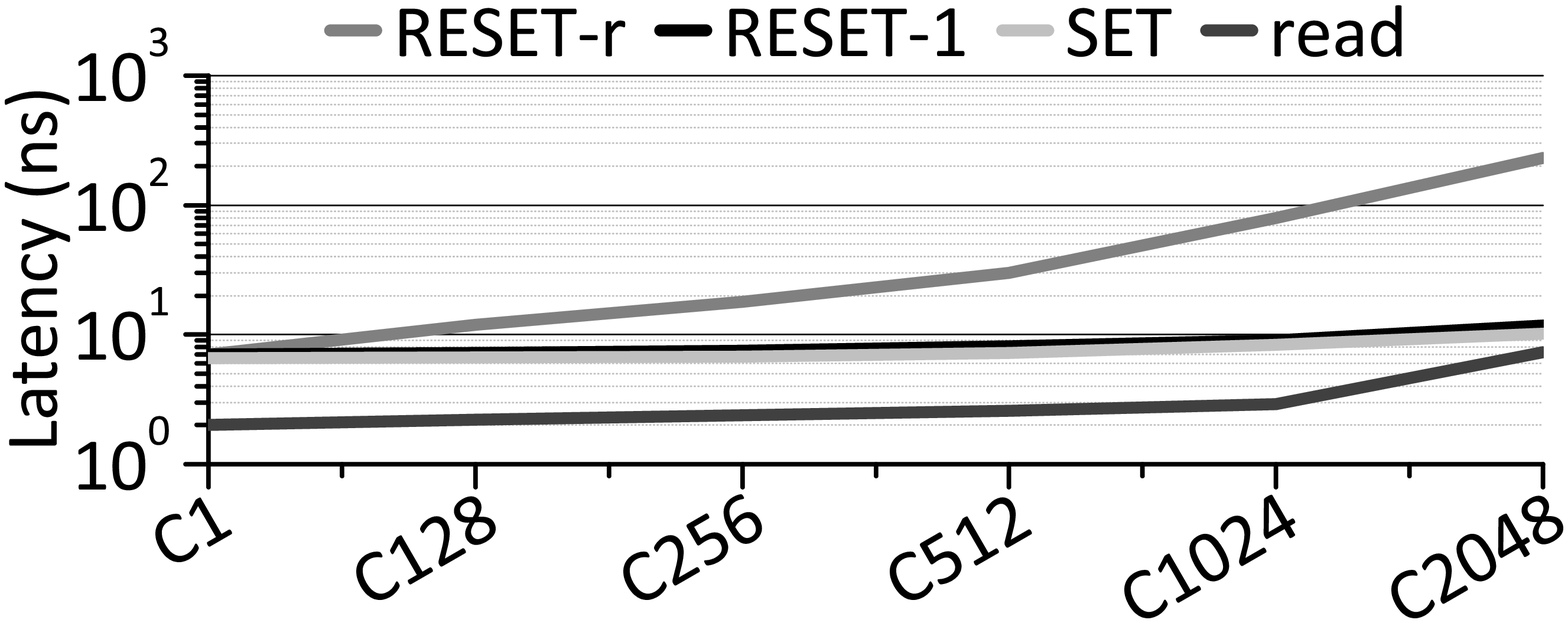}
   \captionof{figure}{The RHU latency.}
   \label{f:dna_rhu_latency}
\end{minipage}%
\begin{minipage}{.65\columnwidth}
   \centering
   \includegraphics[width=\columnwidth]{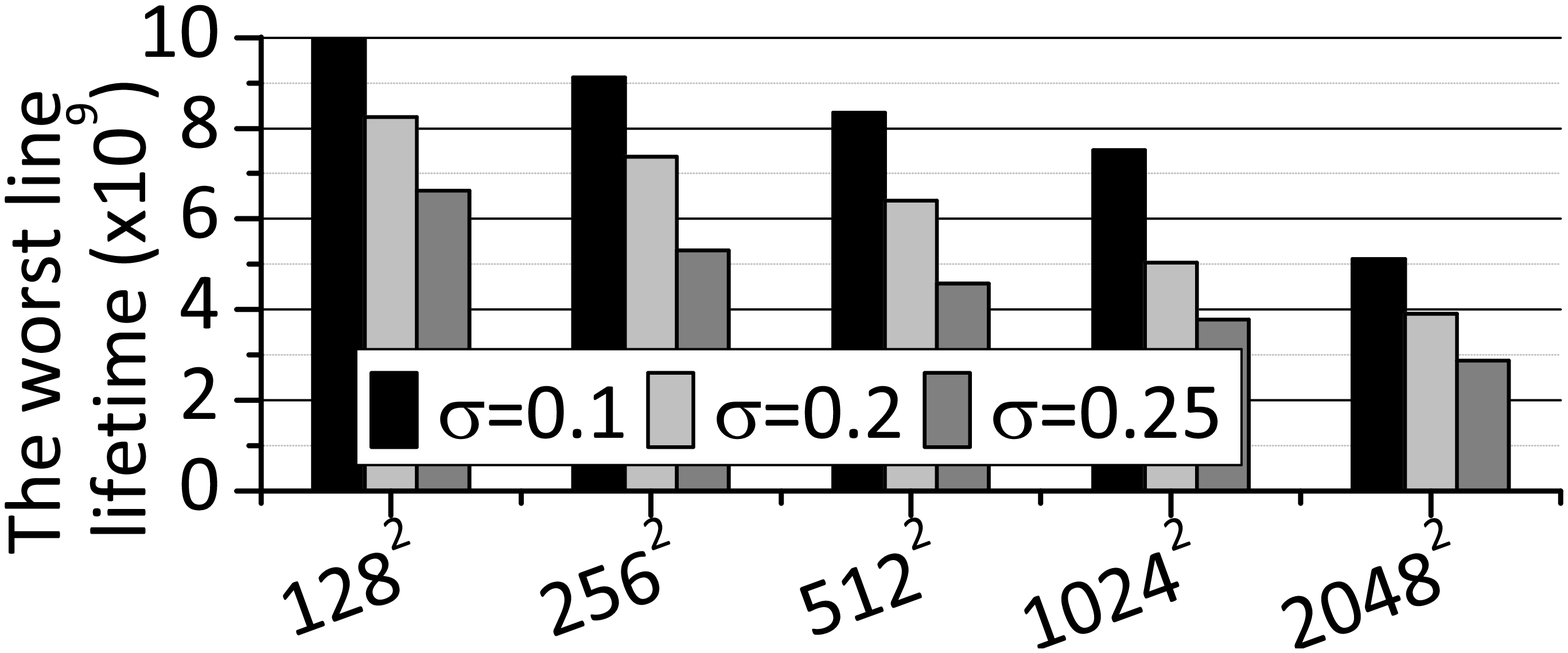}
   \captionof{figure}{The average RHU lifetime.}
   \label{f:dna_rhu_end}
\end{minipage}%
\begin{minipage}{.65\columnwidth}
   \centering
   \includegraphics[width=\columnwidth]{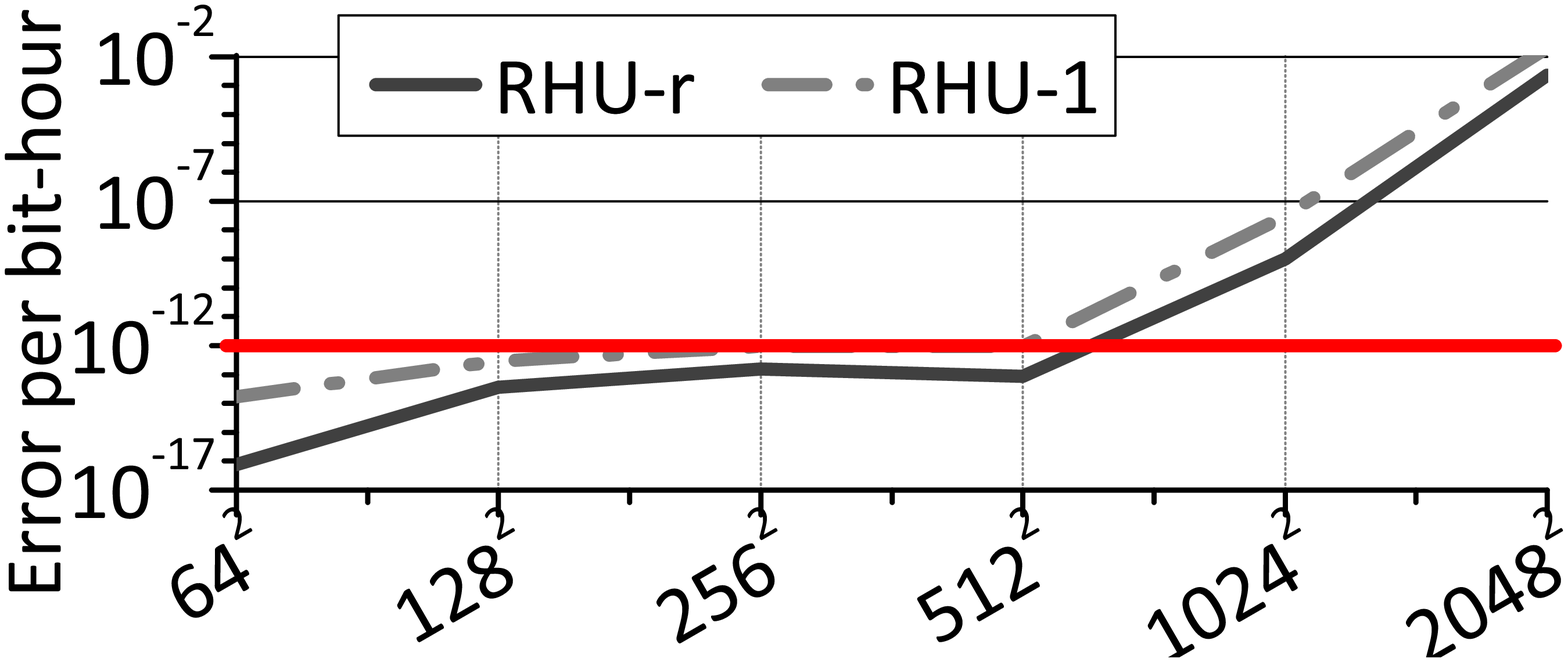}
   \captionof{figure}{The RHU accuracy.}
   \label{f:dna_rhu_acc}
\end{minipage}
\vspace{-0.25in}
\end{figure*}

\subsubsection{The Incompatibility of ReRAM and Logic Processes}

The ReRAM fabrication process and the logic fabrication process are different and often incompatible~\cite{Ito:IMW2018}. It is difficult to fabricate circuits with a ReRAM process, or ReRAM with a logic process. The gates in a ReRAM process are fully optimized for density and low leakage power by sacrificing performance. The ReRAM process often has $<5$ metal layers, while the logic process has $>12$ metal layers. The circuits fabricated in a ReRAM process suffer from much higher interconnect overhead. On the contrary, building ReRAM cells in a logic process creates embedded ReRAM~\cite{Lee:ASSCC2017}, which increases the cell size to $\mathbf{44F^2}$. Therefore, combining ReRAMs and logic circuits results in low area efficiency and high power consumption. Our FindeR aims to process the kernel of FM-Index searches without CMOS logic circuits and by ReRAM arrays only. 

\section{FindeR}
\label{s:fm_index_acc}

As Figure~\ref{f:fm_index_all}(c) exhibits, the bottleneck of an FM-Index backward search of an $m$-bp read lies in the \textit{LFM} function calculating $Occ(s,i)$, since the LFM function is invoked by both $low$ and $high$ for $m$ times. The function of \textit{LFM} counts the number of times the symbol $s$ appears in a $d$-bp BWT bucket for the FM-Index. We transform a symbol counting operation to a Hamming distance (HD) calculation. Counting the symbol $s$ in a $d$-bp BWT bucket can be computed by subtracting $hd$ from $d$, where $hd$ is the HD between the $d$-bp BWT bucket and a $d$-bp sequence where all symbols are $s$. We propose a reliable, fast and energy-efficient ReRAM-based HD unit (RHU) to accelerate HD computations. To avoid introducing extra CMOS logic gates, we present a ReRAM-based lookup table (LUT)-based adder to subtract $hd$ from $d$ inside a ReRAM chip. We further architect a full-fledged pipeline with lightweight scheduling and a system library to improve the FM-Index search throughput.

\subsection{ReRAM-based Hamming Distance Unit}
\label{s:fmindex_rhu_unit}

\subsubsection{An RHU for the FM-Index}
\begin{figure}[htbp!]
\vspace{-0.15in}
\centering
\includegraphics[width=3.3in]{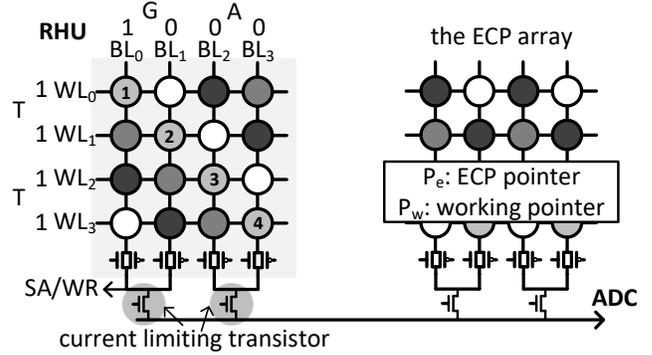}
\vspace{-0.05in}
\caption{An RHU in a ReRAM bank.}
\label{f:dna_hd_unit}
\vspace{-0.15in}
\end{figure}

Figure~\ref{f:dna_hd_unit} illustrates an RHU. In a $4\times4$ ReRAM array, only $4$ main diagonal cells are used to calculate the HD between a $2$-bp BWT bucket and a $2$-bp read ``$ss$'', where $s$ can be $A$, $C$, $G$ and $T$. All the other cells are in HRS. But they cannot be SET or RESET, since they are not FORMed and thus have no filament. There are three steps to compute the HD value between the BWT bucket and the read, e.g., $GA$ and $TT$. First, all diagonal cells in the array are initialized to HRS before an HD calculation. Second, we encode $A$, $C$, $G$ and $T$ by $00$, $01$, $10$ and $11$, and use $0V$ and $1.5V$ to indicate $0$ and $1$. $1.5V$ is the minimal voltage triggering unipolar SETs. But it is not high enough to start FORMING operations~\cite{Wong:IEEE2012}. The corresponding voltage levels indicating the binary encoding of DNA symbols are applied on the WLs and BLs, respectively. For instance, the voltage corresponding to ``$1000$'' ($GA$) is applied on BLs and the voltage corresponding to ``$1111$'' ($TT$) is assigned to WLs. Each cell on the main diagonal line in the array remains in HRS if its WL and BL have the same voltage. The cell \ding{182} is in this case. Otherwise, the cell is switched to LRS. So the cells \ding{183}$\sim$\ding{185} are switched to LRS. One DNA symbol mismatch generates one or two LRS cells. Third, the sensing voltage ($1V$) is applied to all BLs, while all WLs are grounded. A current limiting transistor connected to a pair of BLs passes at most only $1\times$ LRS sensing current even when both cells on two BLs are in LRS. The cells \ding{182} and \ding{183} supply $1\times$ LRS sensing current, while the cells \ding{184} and \ding{185} provide another $1\times$ LRS sensing current, although both of them are in LRS. The summed current representing the HD between $GA$ and $TT$ is measured and translated to a digital HD value ($hd$) by an analog-to-digital converter (ADC) shared by multiple RHUs. $hd$ is 2 in this example. Instead of counting the symbol $s$, the HD indicates the number of symbols different from $s$ in the BWT bucket. The number of symbol $s$ in a $d$-bp BWT bucket is calculated by $d-hd$. Therefore, the number of $T$ in $GA$ is $2-2=0$. An RHU is a normal ReRAM array that can be accessed by both SAs and WDs in a chip.

\subsubsection{The Fabricated RHU Demonstration}

The basic function of a unipolar switching RHU has been examined in~\cite{Ning:SR2017}. However, the existing RHU demonstration relies on one line, but not an array, to perform HD calculations, since it cannot isolate the other cells in the array by skipping their FORMING operations. The existing RHU cannot be used to compute the HD between two genomic reads, since each symbol in a read is encoded by two bits. Without the current limiting transistor, an RHU will produce wrong HD values, since two LRS cells may be generated by a mismatch between two symbols. Moreover, an RHU fails within minutes when constantly computing HD values, due to the short ReRAM cell endurance. We propose wear-leveling and error-correcting schemes to provide RHUs sufficiently long lifetime. We shorten the initialization, HD calculation and popcount latencies of RHU and enhance the current accumulation accuracy of RHU during popcounts by tuning the RHU array size.

\begin{figure*}[t!]
\centering
\subfigure[Area.]{
   \includegraphics[width=0.48\columnwidth]{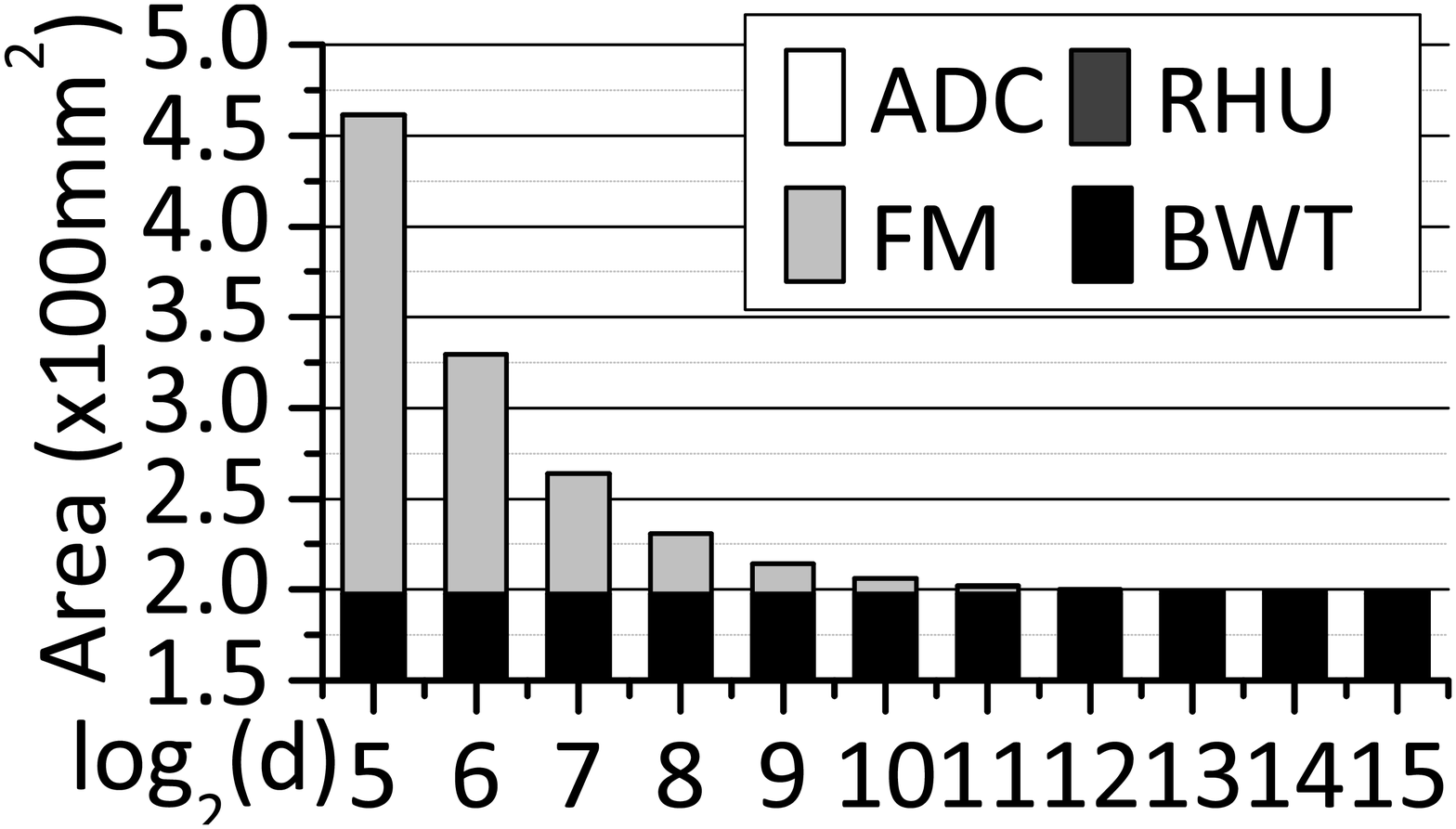}
   \label{f:dna_rhu_area}
}
\hspace{-0.1in}
\subfigure[Power.]{
   \includegraphics[width=0.48\columnwidth]{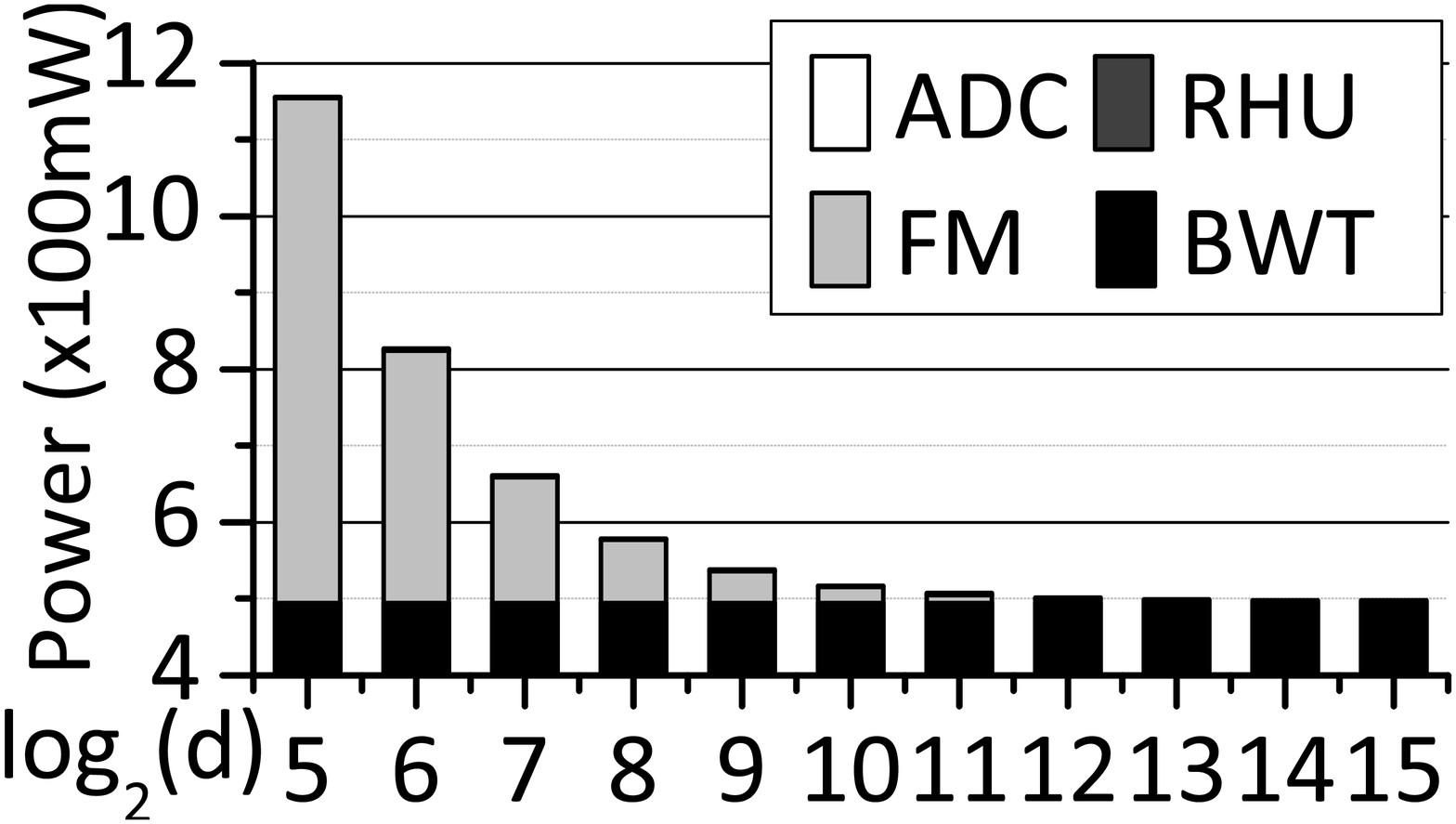}
   \label{f:dna_rhu_power}
}
\hspace{-0.1in}
\subfigure[Energy.]{
   \includegraphics[width=0.48\columnwidth]{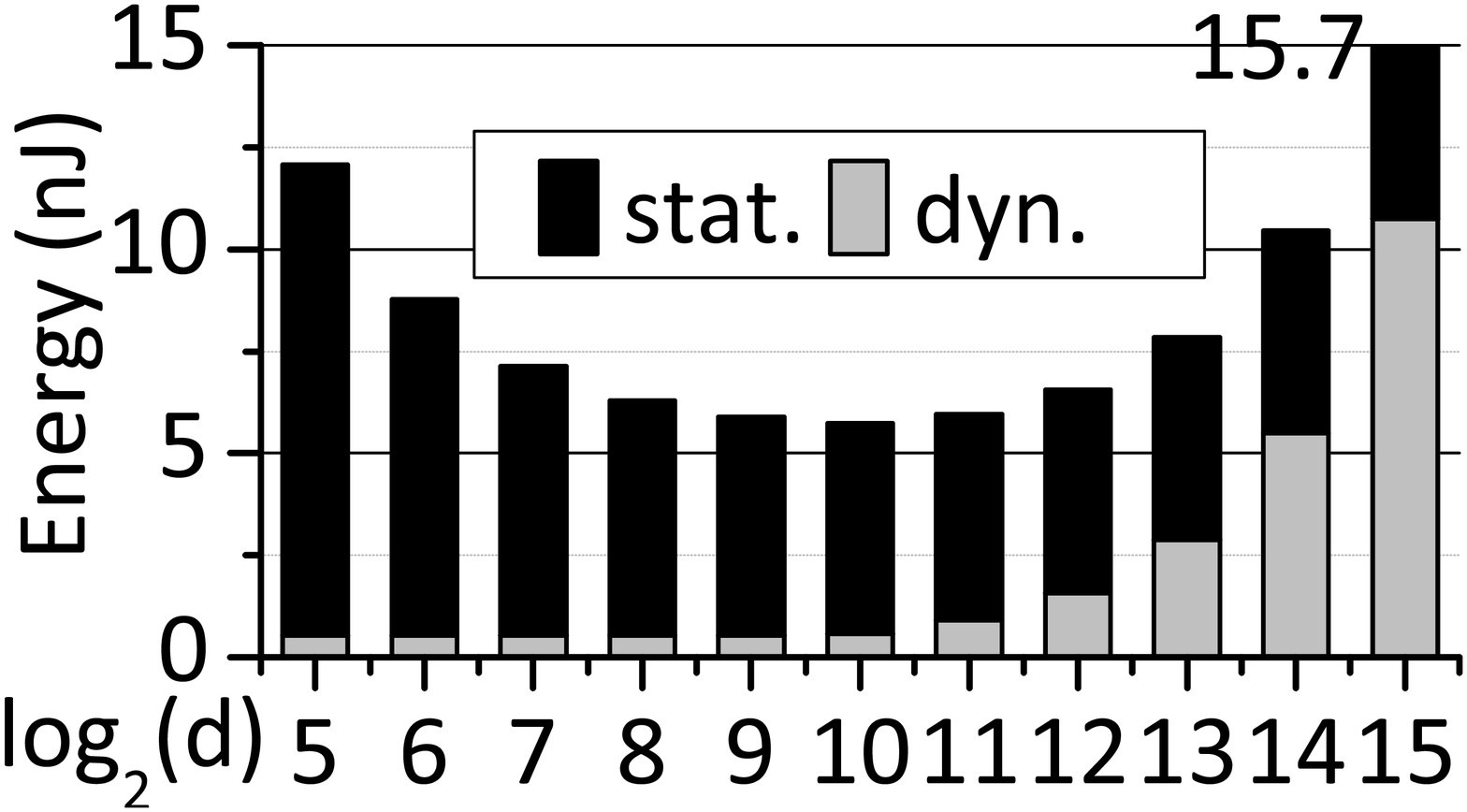}
	 \label{f:dna_rhu_dynen}
}
\hspace{-0.1in}
\subfigure[Lifetime.]{
   \includegraphics[width=0.48\columnwidth]{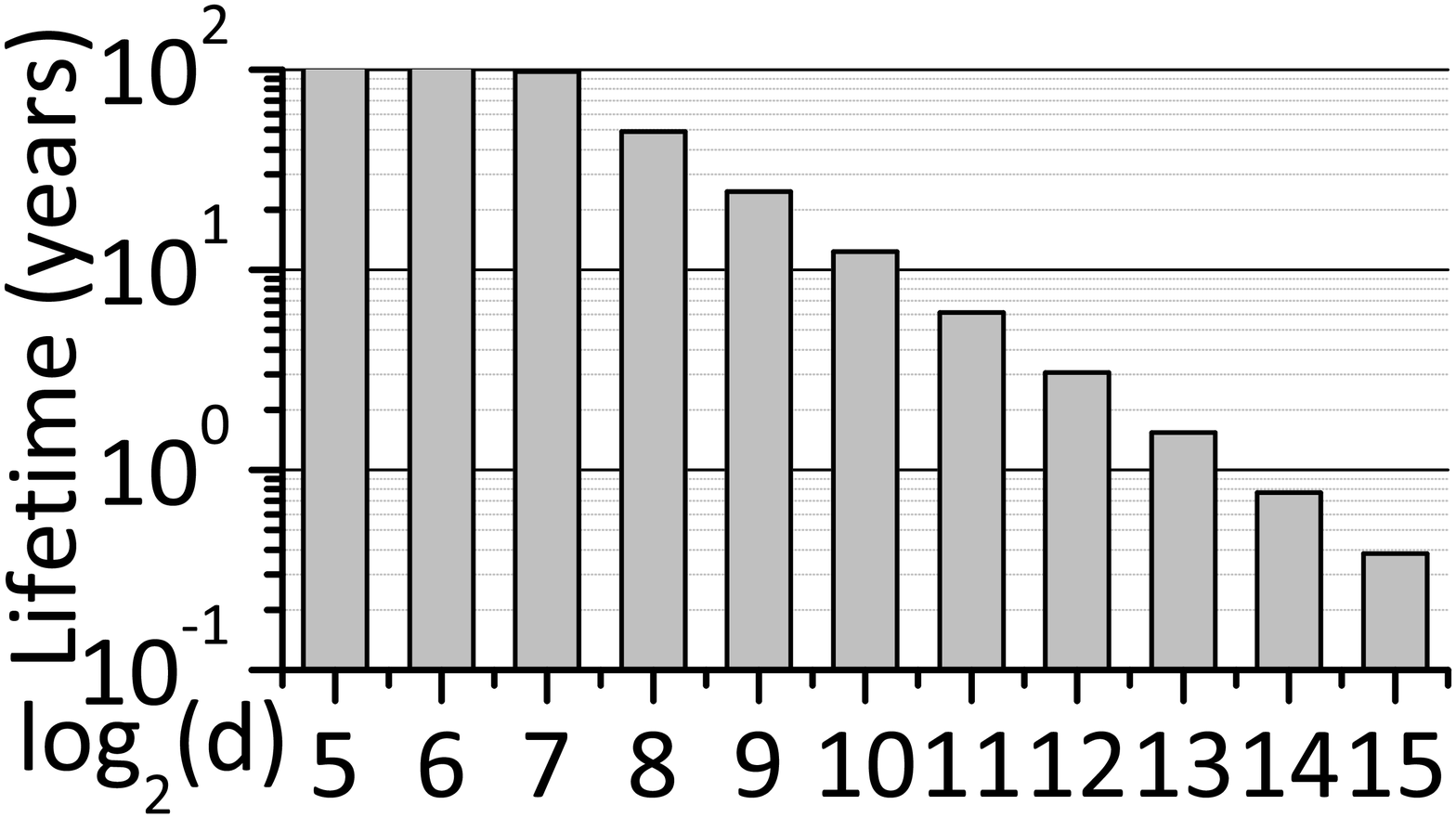}
	 \label{f:dna_rhu_end2}
}
\vspace{-0.1in}
\caption{The design space exploration of RHU with varying BWT bucket widths ($d$). (\textbf{All Y-axes are $log_2(d)$})}
\label{f:design_space_all}
\vspace{-0.3in}
\end{figure*}

\subsubsection{The RHU Latency}

An RHU HD computation involves a RESET initialization, a SET and a read operation. The latencies of cells at different positions on a BL are shown in Figure~\ref{f:dna_rhu_latency}. We adopted the ReRAM cell and array models from~\cite{Mao:JETC2016}. We define $Cell_1$ ($C1$) as the cell closest to the sense amplifiers (SAs) and write drivers (WDs), while $Cell_{2048}$ ($C2048$) is the cell farthest from the WDs on a BL. The SET latency is insensitive to cell position on the BL and only slightly increases for cells far from the WDs. In contrast, as prior works~\cite{Xu:HPCA2015,Wen:ICCAD2017,Mao:JETC2016} indicate, the RESET latency is exponentially prolonged by the voltage drop introduced by the BL resistance and sneak path current. The sneak path is heavily influenced by the data pattern in a ReRAM array~\cite{Wen:ICCAD2017}. If more cells on a BL or WL are in LRS, a larger sneak path current emerges on the BL or WL, and thus the RESET latency is longer. With a random data pattern (the one and zero ratio: 1-to-1), the cells far from the WDs on the BL suffer from large RESET latencies (\textit{RESET-r}). However, the data pattern (\textit{RESET-1}) where only main diagonal line cells are in LRS and all other cells are in HRS does not significantly increase the RESET latency even for cells far from the WDs, since for each pair of connected BL and WL there is only one LRS cell. A RESET on $Cell_{1024}$ on the BL takes only $9ns$ in an RHU. A long BL also slows down a read operation because of the larger RC delay. Since every HD calculation resets the RHU and sets the main diagonal line cells, the RHU can use a high voltage to accelerate the read operation without worrying that the high voltage may disturb cells. We adopted $1.0V$ as the read voltage during an HD calculation.

\subsubsection{The RHU Endurance}

Each ReRAM cell normally tolerates $10^{10}$ writes~\cite{Wong:IEEE2012}, but some weak cells last for only $10^5\sim10^6$ writes because of process variation. Since the RHU uses only the white main diagonal line in Figure~\ref{f:dna_hd_unit}, it fails within minutes when constantly computing HDs. To prolong the endurance of an RHU, we propose a wear leveling scheme to use a pair of other diagonal lines, e.g., two gray, light gray or dark gray lines, for an HD calculation in Figure~\ref{f:dna_hd_unit}. For each 100K HD calculations, we perform BREAKING operations to eliminate filaments in all cells of the current diagonal line pair, and conduct FORMING operations on another diagonal line pair. We use a pointer $P_{w}$ to indicate which diagonal line pair is working in an RHU. $P_{w}$ costs $log_2(w)$ bits, where $w$ is the array width. To further improve the RHU lifetime, we adopted six Error Correcting Pointers (ECPs)~\cite{Schechter:ISCA2010} for the working diagonal line pair. If the white main diagonal line is working and its first cell fails (Figure~\ref{f:dna_hd_unit}), we use a pointer $P_{e}$ to record the broken cell position in the diagonal line and two BLs in the ECP array to replace $BL_0$ and $BL_1$. With the help from error detecting units~\cite{Schechter:ISCA2010} in ReRAM chips, the RHU can reuse the same 6 $P_{e}$s for all its diagonal line pairs. Six $P_{e}$s occupy $6log_2(w)$ bits in a dedicated array. For instance, for a $1024\times1024$ array, we have 12 BLs as the BL replacement in the ECP array and 70 bits for $P_{w}$ and $P_{e}$s in a dedicated array. We modeled the cell endurance variation among 4GB cells by the {\it normal} distribution, where $\mu=10^{10}$ and $\sigma=0.1\sim 0.25$~\cite{Schechter:ISCA2010}. The RHU lifetime is decided by its weakest diagonal line endurance with the protection of 6 ECPs and our wear leveling. The average RHU lifetime with varying array sizes is shown in Figure~\ref{f:dna_rhu_end}. On average, a larger RHU array has less ECPs per cell, so its lifetime is shorter. Even with $\sigma=0.25$, the average lifetime of a $2048\times2048$ ($2048^2$) RHU is $2.87\times 10^9$ writes. We will show the endurance of the whole FindeR PIM in Section~\ref{s:dse}.

\subsubsection{The RHU Accuracy}

Based on Kirchoff's Law, an RHU sums currents passing through the cells in a diagonal line pair as the popcount result. We need to study the accuracy of current accumulation, which is influenced by the cell resistance variation, the wire resistance and the random telegraph noise (RTN)~\cite{Lee:EDL2011}. We adopted the ReRAM cell and array models from~\cite{Mao:JETC2016}. The LRS and HRS resistance variations are modeled through the {\it normal} distribution, where for LRS $\mu=2k\Omega$, HRS $\mu=2M\Omega$ and $\sigma=0.03$~\cite{Park:IMW2013}. We modeled the RTN by adding 2\% and 3\% of binary noises to the LRS and HRS resistances~\cite{Lee:EDL2011}. We generated 4GB cells to study the accuracy of current accumulation. We produced random data (the one and zero ratio: 1-to-1) as the normal data pattern (\textit{RHU-r}) and used \textit{all-1}s as the worst case data pattern (\textit{RHU-1}) for the cells in each diagonal line pair. Figure~\ref{f:dna_rhu_acc} exhibits the RHU current accumulation accuracy. RHU counts the number of LRS cells in a diagonal line pair. Unlike deep learning applications, FM-Index searches for exact pattern matching in genome sequences are more vulnerable to errors resulting in wrong pattern matching. The RHU should have similar reliability to a CMOS-based counterpart with SRAM registers. So we used the SRAM soft error rate, $10^{-12}$ error per bit-hour~\cite{Tezz:SOFT2004}, as an acceptable error rate for the current accumulation (popcount) of RHU. The RHU can guarantee such low error rate with the worst \textit{all-1}s data pattern (\textit{RHU-1}) when counting all cells along a diagonal line pair even in a $512\times512$ array ($512^2$). This is because all of the cells of the RHU are in HRS except the cells in the current diagonal line pair. The structure of two BLs sharing one current-limiting transistor also alleviates the negative impact of the cell resistance variation and RTN.

\begin{figure}[t!]
\centering
\includegraphics[width=3in]{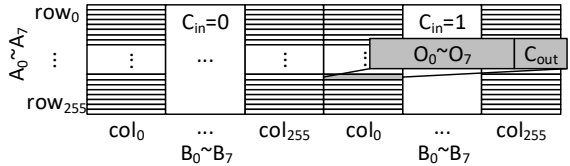}
\vspace{-0.1in}
\caption{A ReRAM-based 8-bit LUT adder.}
\label{f:dna_lut_adder}
\vspace{-0.3in}
\end{figure}

\subsection{ReRAM Lookup Table-based Adder}

To avoid adding extra CMOS logic into a ReRAM chip, we propose a ReRAM lookup table (LUT)-based adder to calculate the sum of a marker and an RHU output. Figure~\ref{f:dna_lut_adder} illustrates a ReRAM-based 8-bit LUT-based adder with three inputs $A_7\sim A_0$, $B_7\sim B_0$ and $C_{in}$. Through $A_7\sim A_0$, the adder activates the corresponding WL. It uses $B_7\sim B_0$ and $C_{in}$ to select BLs to read the 9-bit result, including an 8-bit sum $O_7\sim O_0$ and a 1-bit $C_{cout}$. The 8-bit LUT-based adder occupies 0.14MB cells. Because the markers of the FM-Index are 32-bit, we conduct 32-bit additions by four lookups on the 8-bit adder. We first send the 8 least significant bits (LSBs), $LSB_7\sim LSB_0$, of the marker and the RHU output to the adder with $C_{in}=0$. After the first lookup is done, the next 8 LSBs of two operands are assigned to the adder with $C_{in}=C_{cout'}$, where $C_{cout'}$ is the output of $C_{cout}$ from the previous lookup. The lookups continue until all bits of two operands have been processed. To sum a marker and an RHU output $hd$, the FM-Index marker is modified to $marker+d$, where $d$ is the FM-Index bucket width. And the LUT-based adder actually stores the values of $A_7\sim A_0$ minus $B_7\sim B_0$ to compute $marker+d-hd$.

\begin{figure}[htbp!]
\vspace{-0.2in}
\centering
\includegraphics[width=3.3in]{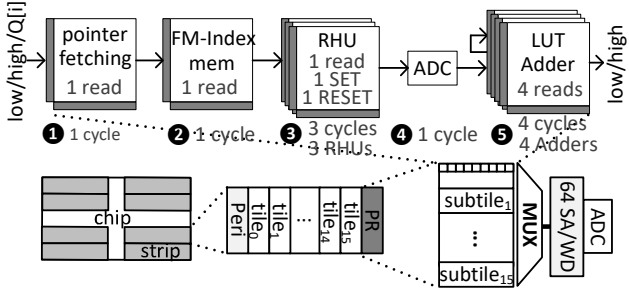}
\vspace{-0.1in}
\caption{The FindeR pipeline design.}
\label{f:dna_pipeline_des}
\vspace{-0.2in}
\end{figure}

\subsection{Pipeline Design}

\subsubsection{Pipeline Details}

FM-Index searches need to invoke the $LPM$ function to update $low$ and $high$. We propose a pipelined FindeR in Figure~\ref{f:dna_pipeline_des} to increase the throughput of the $LPM$ function. The pipeline includes five stages: pointer fetching, FM-Index memory reading, RHU, ADC and adder. \ding{182} During the pointer fetching stage, the FM-Index searches read the working array pointer, the working diagonal line pointer ($P_w$), and ECPs ($P_e$s) from their dedicated arrays. \ding{183} The FM-Index then fetches a $d$-symbol BWT bucket ($2d$-bit vector) with markers from FM-Index data arrays. \ding{184} The BWT (bit vector) and the read symbol are processed by an RHU specified by the pointers retrieved during \ding{182}. The summed current of an RHU is sensed and held by a transimpedance amplifier (TIA). Besides a SET and a read, for each HD calculation, the RHU also needs to conduct a RESET initialization applying $V_{RESET}=1V$ to all BLs and $0V$ to WLs without compliance current. So we require three RHUs to create an analog HD value each pipeline cycle. \ding{185} The current is converted to a digital value $hd$ by a 128MHz 8-bit ADC~\cite{Kull:ISSCC2013}. \ding{186} The value of the $d-hd+marker$ for the FM-Index is calculated by looking up the adder. One addition consists of four lookups, each of which can be done within one pipeline cycle. So we integrate four adders into our pipeline to produce a result every cycle. We set one pipeline cycle as $10ns$, since as Section~\ref{s:fmindex_rhu_unit} explains, each $1024\times1024$ RHU read, SET, RESET and ADC conversion can be completed within $10ns$. In this way, the pipeline can be operated at $100MHz$. The function \textit{LFM} of the FM-Index takes $90ns$ (i.e., pointer $10ns$, data $10ns$, RHU $20ns$, ADC $10ns$ and adder $40ns$) to calculate a $low$ or $high$ by this pipeline design.

\subsubsection{Enabling Strip-Level Parallelism}

To process big genomic data, we assume a high density ReRAM-based main memory system~\cite{Xu:HPCA2015}, where eight ReRAM chips are interleaved to form eight banks through an NVDIMM. One FindeR pipeline is integrated into each bank, so that each pipeline can operate independently. In our ReRAM-based main memory, one bank can only serve one 64B read or write at one time. Each bank has eight strips, each of which is in a chip. Each strip is equipped with 64 SAs and 64 WDs as shown in Figure~\ref{f:dna_pipeline_des}. To build a pipeline design inside each bank, each strip has to serve a smaller size access independently. We adopted the low overhead two-dimensional bank subdivision~\cite{Poremba:DAC2016} to enable multiple simultaneous small size accesses in a bank. We add four 0.14MB arrays, each of which has 9 sense amplifiers and 9 write drivers, as the LUT adders in each bank. With the exception of adder arrays, FindeR requires only another 5 independent strips to run the pipeline in a bank.

\subsection{FindeR PIM}
\label{s:dse}

\subsubsection{Design Space Exploration}

Figure~\ref{f:design_space_all} highlights the design space exploration of the RHU for the FM-Index with varying bucket widths ($log_2(d)$s). To achieve high memory density, we used $\mathbf{1024\times1024}$ arrays to build FindeR, since the SET, RESET and read operations in such large arrays can be completed within a pipeline cycle \textbf{10ns}. With increasing bucket size, the FM-Index storage is substantially reduced. The area and power for FM-Index, shown in Figure~\ref{f:dna_rhu_area} and~\ref{f:dna_rhu_power} decrease with a larger $log_2(d)$, since the FM-Index occupies less arrays. However, the $LFM$ function has to count a symbol in a larger BWT bucket when $d$ increases. The RHU has to SET more cells to complete an HD calculation within the one pipeline cycle. As Figure~\ref{f:dna_rhu_end2} exhibits, the lifetime of 16K $1024\times1024$ RHU arrays (\textbf{2GB}) decreases when $d$ increases, because more writes occur in an RHU array during each HD calculation. With $log_2(d)=10$ and large process variation ($\sigma=0.25$), a FindeR consisting of 2GB RHU arrays can still stand for more than \textbf{10 years} even when constantly computing HD values. More concurrent SETs also substantially boost the RHU dynamic energy consumption in Figure~\ref{f:dna_rhu_dynen}. Considering that the RHU current accumulation accuracy is acceptable when counting $\leq 512$ cells in a diagonal line, we conservatively selected $\mathbf{d=128}$ to balance the area, power, endurance and current accumulation accuracy for the RHU.

\subsubsection{Search Iteration Scheduling}

A slave memory controller (SMC)~\cite{Ham:HPCA2013,Fang:PACT2011} is deployed on the NVDIMM to translate read/write requests from CPUs to ReRAM device-specific commands (CMDs) that can be processed by banks. For the FM-Index, the SMC issues requests with read symbols to perform backward search iterations. The request addresses are calculated through $low$, $high$ and the bucket width $d$. After receiving the new $low$ and $high$, the SMC schedules the next search iteration of the FM-Index as follows: when $low<high$, the $low$ and $high$ will be returned to the on-chip master MC. A backward search is done. When $low>high$, the SMC creates requests, decodes their bank numbers by their address, and issues them into bank queues. When $low==high$, two requests for $low$ and $high$ ask for the same BWT bucket. The SMC can coalesce these two requests. It issues two requests into the same bank queue and sets an indication bit in the $low$ request, so that the bucket data can be kept in sense amplifiers of the bank. The $high$ request reads the data directly from SAs. To maintain the pipeline frequency, the coalescing does not reduce the $high$ request latency, but it minimizes the read energy. Based on our experiments, during short read alignment of human genome, 85\% of symbols in a 100-bp read can coalesce their $low$ and $high$ requests on average.

\begin{figure}[htbp!]
\vspace{-0.15in}
\centering
\includegraphics[width=\linewidth]{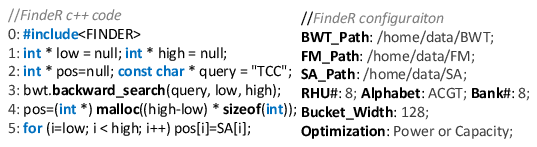}
\vspace{-0.2in}
\caption{The system support of FindeR.}
\label{f:dna_rhu_system}
\vspace{-0.15in}
\end{figure}

\subsubsection{System Support}

To run FindeR as a PIM accelerator in a computing system with processors, the FindeR integration stack includes three parts: configuration file, programming support and run-time library. Figure~\ref{f:dna_rhu_system} describes a library configuration example and a code snippet of FindeR. In the \textit{configuration} stage, programmers can easily specify the basic FindeR parameters, e.g., the BWT and FM-Index files, alphabet, FM-Index bucket width, bank number and RHU number, in the configuration file. We assume the BWT construction of the reference genomes and read pools are done in the cloud~\cite{Liu:TCBC2016,Liu:BIOIN2016}, so that we can perform trillions of backward searches on them during all steps of genome analysis. At the beginning of compiling, the files of the BWT and FM-Index are copied into ReRAM chips and the other parameters are written into the SMC on the NVDIMM. The \textit{programmers} can also enable FM-Index optimizations for FindeR in the configuration file by specifying optimization goals such as power or capacity. Based on the existing parameters, e.g., bank number, RHU number and alphabet, the FindeR library calculates the FM-Index bucket width to achieve the optimization goal. With the FindeR feedback and hint, programmers may re-compute the FM-Index for existing BWTs to attain smaller power and memory capacity locally or in the cloud. During programming, FindeR provides APIs for programmers to allow fast FM-Index-based EPM operations. When the compiled code \textit{execute}s, FindeR accelerates FM-Index backward searches in ReRAMs and returns $low$ and $high$ pointers to the CPU. To retrieve EPMs, the CPU fetches the SAs allocated in idle banks by $low$ and $high$ pointers.

\subsubsection{Hardware Overhead}

\begin{figure}[htbp!]
\vspace{-0.1in}
\centering
\subfigure[Area and Power.]{
   \includegraphics[width=3.3in]{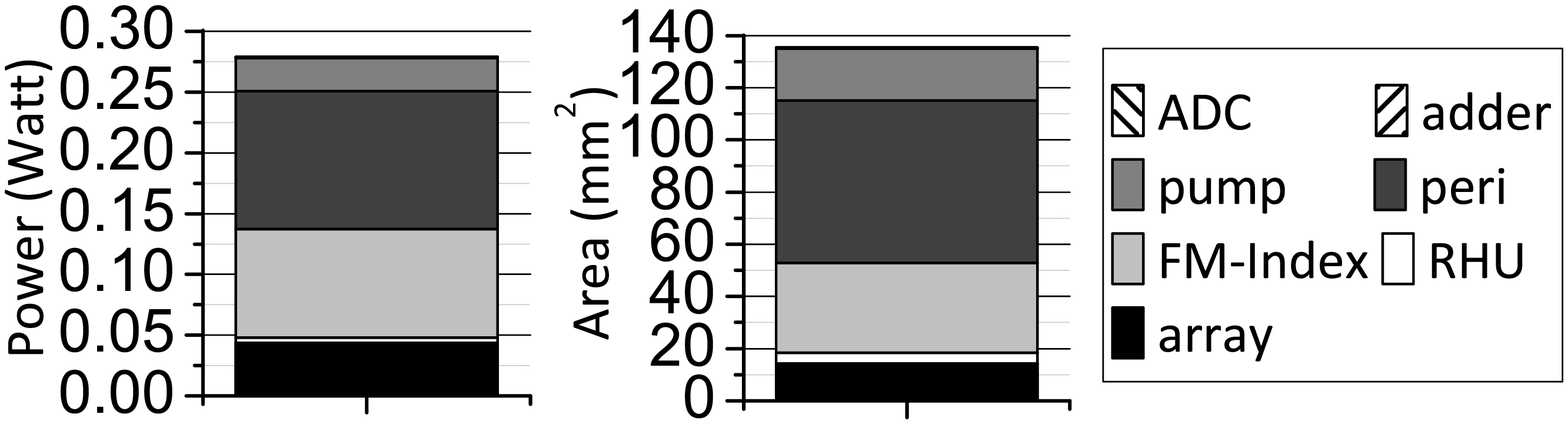}
	 \label{f:dna_all_result}
}\\
\vspace{-0.1in}
\subfigure[Energy.]{
   \includegraphics[width=3.3in]{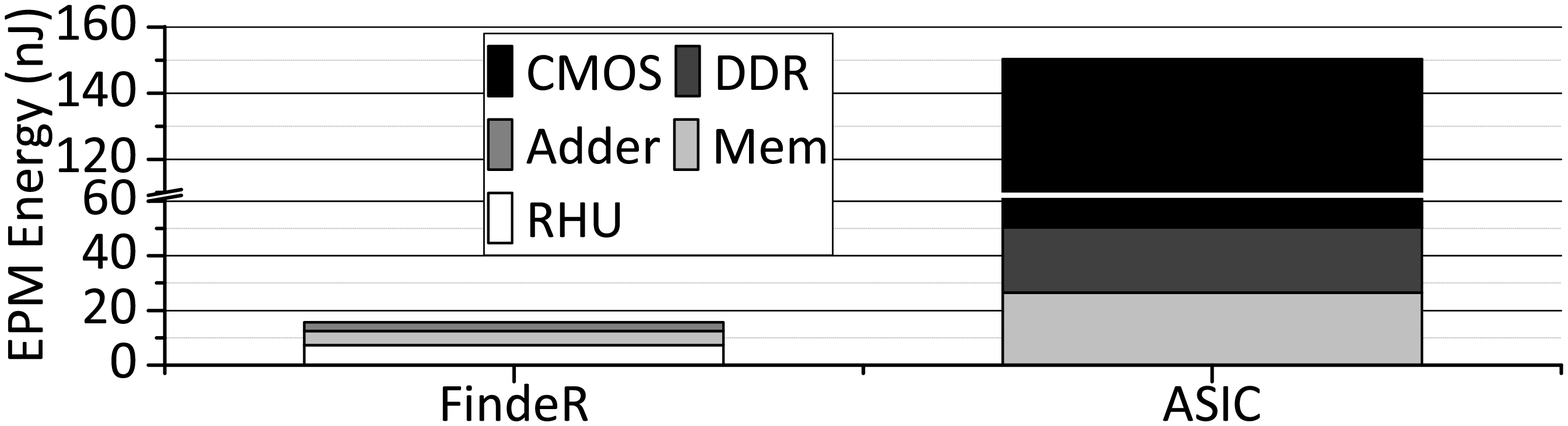}
	 \label{f:dna_energy_result}
}
\vspace{-0.1in}
\caption{The hardware overhead of FindeR.}
\label{f:dna_all_result000}
\vspace{-0.1in}
\end{figure}

We modeled the power, energy, latency and area of FindeR by NVSim~\cite{Li:ICCAD2016} calibrated by the latest NVDIMM ReRAM power model~\cite{Choi:IMW2018}. FindeR consists of 8 banks, each of which supports a 64B access and is interleaved across 8 chips. When the bucket width is $d=128$, FindeR stores the bi-directional FM-Index with 2.6GB ReRAM arrays in each bank. As Figure~\ref{f:dna_all_result} shows, one 4GB ReRAM bank consumes $135mm^2$ area and $0.279W$ power at $32nm$. It spends $7.1nJ$ energy during each pipeline cycle. We assumed the FindeR NVDIMM has 8 4GB ReRAM banks (32GB) in the case that programmers may change $d$. Programmers can enable multiple banks to process FM-Index searches simultaneously. The idle arrays in each bank can be used for ECPs and main memory~\cite{Chi:ISCA2016}. The bank subdivision~\cite{Poremba:DAC2016} enabling independent strip accesses increases the ReRAM chip area by 5\% and the power consumption by 3.2\%. Four 8-bit LUT adders in each bank cost 0.56MB cells and require 36 SAs. An ADC is an essential component in ReRAM chips~\cite{Park:IMW2013} for reference cell calibrations. We adopted an 8-bit ADC~\cite{Kull:ISSCC2013} in each bank and lowered its frequency to 128M Sample/sec to save power. We synthesized the ADC and NVDIMM scheduling logic by the Cadence design compiler with 32nm PTM process technology. The ADC costs $0.0012mm^2$ and $0.2mW$, while the scheduling logic consumes $0.00014mm^2$ and $0.001mW$. Unlike prior ReRAM-based convolutional neural network accelerators~\cite{Shafiee:ISCA2016,Song:HPCA2017}, \textbf{the overhead of FindeR ADC power and area is trivial}, since it only handles binary HD computations but not floating point arithmetic. We show the energy consumption of an EPM in FindeR in Figure~\ref{f:dna_energy_result}. Compared to an ASIC design~\cite{Wu:ITBCS2017} with 1.3GB DDR4 DRAM, FindeR reduces the energy per EPM by 89.6\%. The DRAM has to open a 2KB row in each read, so even the energy consumed by DRAM arrays in an ASIC-based EPM is larger than the energy of an entire FindeR-based EPM.

\section{Experimental Methodology}
\label{s:eandm} 

\subsection{Simulation and evaluation}

We modified the NVM timing simulator NVMain~\cite{Poremba:CAL2015} to model the micro-architecture of FindeR. We implemented all pipeline details including pointer and FM-Index arrays, RHUs, ADCs and LUT adders. To estimate the \textbf{throughput} of various hardware platforms, we used the metric of \textbf{reads per second}. To measure the quality of read alignments, from~\cite{Turakhia:ASPLOS2018}, we adopted the metric of sensitivity shown in Equation~\ref{e:sens} and the metric of precision shown in Equation~\ref{e:precision}:
\begin{equation}
TP/(TP+FN)
\label{e:sens}
\end{equation} 
\begin{equation}
TP/(TP+FP)
\label{e:precision}
\end{equation} 
where $TP$, $FP$ and $FN$ are the number of true positives, false positives and false negatives, respectively. A true positive (for long reads) is a read mapped correctly that should be mapped (within 50-bp of the region~\cite{Turakhia:ASPLOS2018}).

\subsection{Workloads}
To evaluate our FindeR, we adopted several state-of-the-art FM-Index-based genome analysis applications: BWA-MEM \cite{Li:BWAMEM2013} for short and long read alignment, SGA~\cite{Simpson:GR2012} for short and long read assembly, Soap3~\cite{Luo:PLOS2013} for GPU-based short read alignment, ExactWordMatch~\cite{Healy:GEN2003} for genome annotation and a reference-based genome compression algorithm~\cite{Prochazka:DCC2014}. For long read alignment and assembly, we also applied the FM-Index-based error correcting technique, FMLRC~\cite{Wang:BMC2018}, to reduce the number of errors.

\subsection{Datasets}

For genome alignment, annotation and compression, we used the chromosomes 1$\sim$22, $X$ and $Y$, from the latest human genome GRCh38 as the reference genome. To study FindeR on short reads, we used the Illumina platinum NA12878 human dataset~\cite{Eberle:GR2016} (ERR194147\_1) consisting of 0.78G reads of 101-bp length with $50\times$ as the short read alignment input. To estimate the performance of FindeR on long reads, we created long reads (with length of 1K-bp) by PBSIM~\cite{Ono:BIOFOR2012}. The error profiles of long reads~\cite{Turakhia:ASPLOS2018} can be summarized in the format of (name, mismatch, insertion, deletion, total), e.g.,. (PacBio, 1.50\%, 9.02\%, 4.49\%, 15.01\%) and (ONT\_2D, 16.50\%, 5.10\%, 8.40\%, 30.0\%). For \textit{de novo} assembly, we used PBSIM and DWGSim~\cite{Li:BIOFOR2009} to generate long and short reads of \textit{C. elegans} with $30\times$ coverage.

\subsection{Schemes}

To evaluate the speedup of all steps in genome analysis, we selected a 4GHz 8-core Intel Xeon E5-2667 (v2) processor. CPU power is measured by the RAPL Interface. To study the performance of FM-Index searches on various hardware platforms, we further chose a 1.3GHz 3840-cudaCore Nvidia Tesla P40 GPU with 24GB GDDR5, an FPGA implementation~\cite{Arram:TCBB2017}, and a $40nm$ ASIC chip~\cite{Wu:ITBCS2017}. GPU power is collected by the Nvidia visual profiler. To estimate the performance of short read alignment, we selected the ASIC designs including RaceLogic~\cite{Madhavan:ISCA2014} and GenAx~\cite{Fuijiki:ISCA2018}, and ReRAM content address memory (CAM)-based PIM architecture including RADAR~\cite{Huangfu:DAC2018}, Bio-CAM~\cite{Karen:CORR2017} and R-CAM~\cite{Kaplan:MICRO2017}. To evaluate the performance of long read alignment, we compared FindeR against a recent ASIC chip Darwin~\cite{Turakhia:ASPLOS2018}. Since different accelerators are fabricated by various process technologies, we scaled all area and power metrics with $32nm$ technology.

\begin{figure}[t!]
\centering
\begin{minipage}{.5\columnwidth}
   \centering
   \includegraphics[width=\columnwidth]{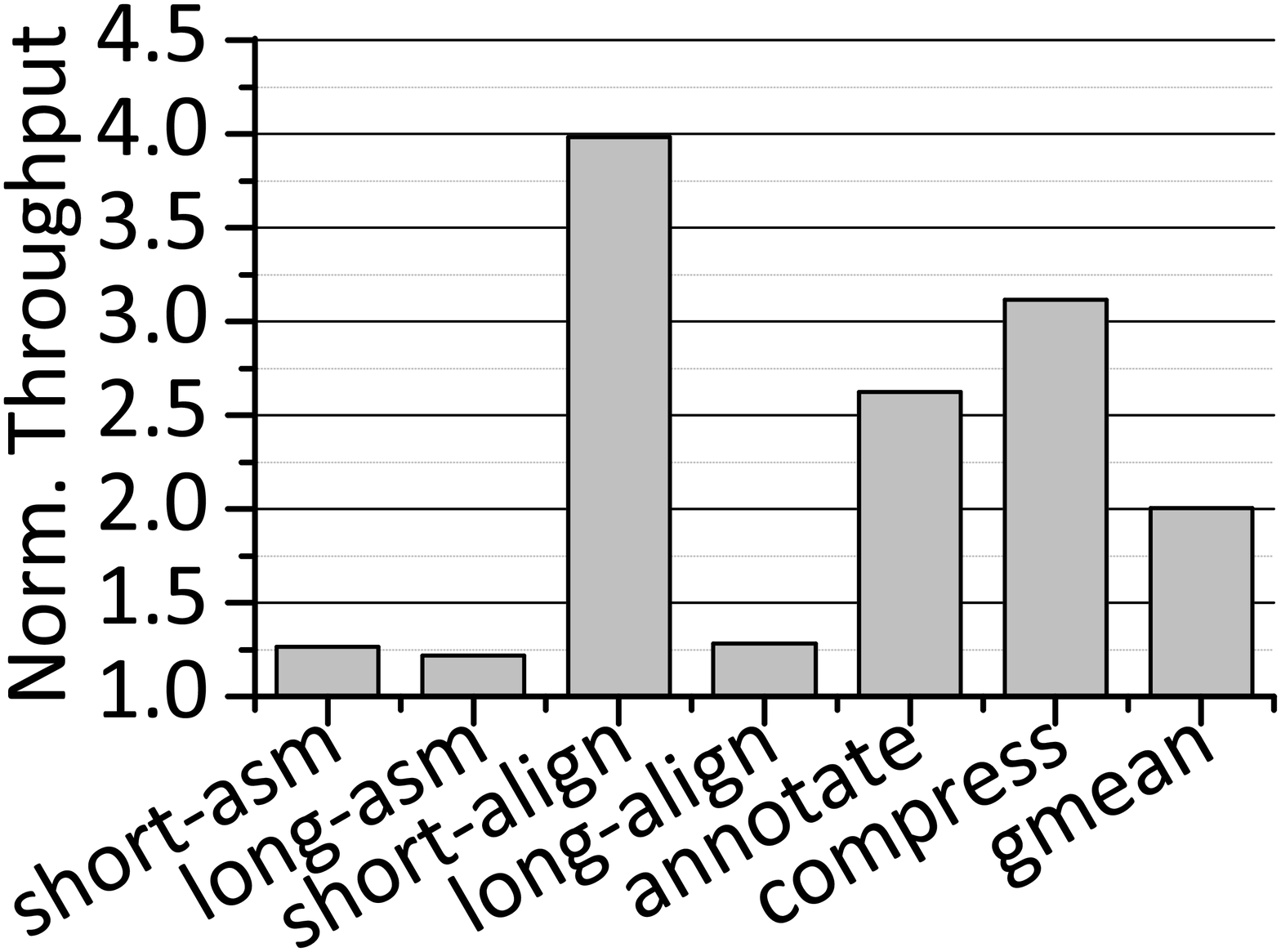}
   \captionof{figure}{Throughput.}
   \label{f:dna_fm_throughput}
\end{minipage}%
\begin{minipage}{.5\columnwidth}
   \centering
   \includegraphics[width=\columnwidth]{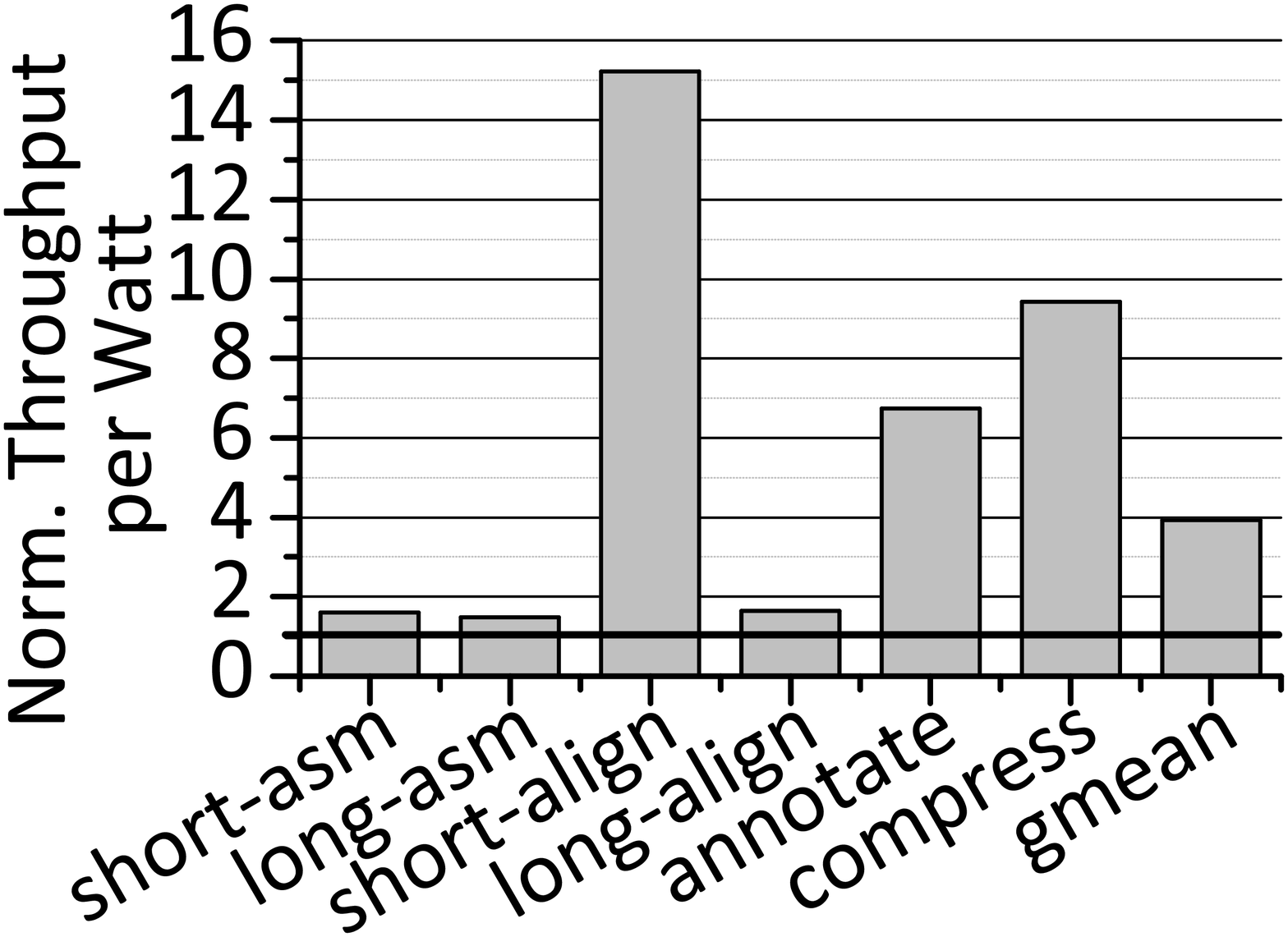}
   \captionof{figure}{Throughput per Watt.}
   \label{f:dna_fm_throughputwatt}
\end{minipage}
\vspace{-0.2in}
\end{figure}

\section{Results and Analysis}
\label{s:reandana}

\subsection{FM-Index in Genome Analysis}

We executed short and long read alignment, short and long read assembly, genome annotation and compression on our CPU baseline which can support all steps in genome analysis. We then moved the execution of FM-Index searches to FindeR. We report the throughput improvement of FindeR over the CPU in Figure~\ref{f:dna_fm_throughput} and the improvement of the throughput per Watt of FindeR over the CPU in Figure~\ref{f:dna_fm_throughputwatt}. During genome annotation and compression, FM-Index search is the most fundamental operation, accounting for $60\%\sim70\%$ of CPU time. For short read alignment, only 0.3\% of reads are processed by the SW algorithm. The bi-directional FM-Index search is used in both SMEM seeding and $\leq2$-mismatch seed extensions for the majority of short reads, so it costs 76\% of CPU time. Therefore, compared to the CPU, FindeR improves the throughput of short read alignment, genome annotation and compression by $2.9\times$, $1.6\times$ and $2.1\times$ respectively. It increases the throughput per Watt of these steps by $14.2\times$, $5.7\times$ and $8.4\times$ respectively over our CPU baseline. For short read assembly, the FM-Index is used only during SMEM seeding. Long read assembly and alignment apply the FM-Index in SMEM seeding constructions and error corrections~\cite{Wang:BMC2018}. But the largest portion ($>50\%$) of execution time during long read assembly and alignment is spent by the seed chaining filter~\cite{Li:BWAMEM2013} that does not invoke FM-Index searches. Compared to the CPU, FindeR improves the throughput of short and long read assembly, and long read alignment by $26.5\%$, $21.9\%$ and $28.1\%$ respectively. It boosts the throughput per Watt of these steps by $59.5\%$, $48.2\%$ and $63.6\%$ respectively over the CPU. On average, compared to the CPU, FindeR improves the throughput and the throughput per Watt of all steps in the genome sequencing by 101\% and 294\%, respectively.

\begin{table}[htbp!]
\setlength{\tabcolsep}{3pt}
\centering
\begin{tabular}{| l |c |c |c |c |c |c |c |}
\hline
                       & CPU     & GPU~\cite{Luo:PLOS2013} & ASIC~\cite{Wu:ITBCS2017} & FPGA~\cite{Arram:TCBB2017} & \textbf{FindeR} \\ \hline\hline
Die size ($mm^{2}$)    & 14.3K   & 1.6K                  	 & 352    	                & 14.8K  	                   & 1.1K            \\ \hline
Main memory(GB)        & 128     & 6                       & 1.3                      & 48                         & 0               \\ \hline
Power(W)               & 130     & 258                     & 3.1                      & 247                        & 9.09            \\ \hline
Throughput             & 68K     & 150K                    & 379K                     & 1.5M                       & \textbf{10.7M}  \\ \hline
Throughput/Watt        & 523     & 581                     & 121K                     & 6.2K                       & \textbf{1.18M} \\ \hline
\end{tabular}
\caption{FM-Index search on various hardware platforms.}
\label{t:hard_perf_v}
\vspace{-0.1in}
\end{table}

\subsection{FM-Index on Various Hardware Platforms}
\label{s:fm_vhp}

We used the SMEM seeding construction of short read alignment to evaluate FindeR and compare it against various hardware platforms, since the CPU-, GPU-, FPGA-, and ASIC-based FM-Index accelerators support the SMEM seeding construction where the FM-Index searches cost $>$97\% of the application execution time~\cite{Chang:FCCM2016}. The FM-Index search throughput, area and power comparison of various hardware platforms is shown in Table~\ref{t:hard_perf_v}. Except our FindeR, all the other FM-Index search accelerators reply on the off-chip main memories to store FM-Index data and reads, and thus waste huge amounts of energy when fetching data from off-chip main memories. Since the CPUs, GPUs and FPGAs adopt software-based multi-step FM-Index optimizations~\cite{Luo:PLOS2013,Arram:TCBB2017} to increase the FM-Index search throughput and fully utilize the hardware resources, they require a large capacity of DRAMs. The BWT of the FM-Index effectively compresses the genome reference data, so it is possible for the ASIC~\cite{Wu:ITBCS2017} to conduct FM-Index backward searches using only 1.3GB DRAM. Compared to the ASIC design, FindeR boosts the throughput and throughput per Watt of FM-Index searches by $28.2\times$ and $9.75\times$.

\begin{table*}[htbp!]
\setlength{\tabcolsep}{3pt}
\centering
\begin{tabular}{| l |c |c |c |c |c |c |}
\hline
\multicolumn{1}{|c|}{}  & \multicolumn{2}{c|}{Hash Table} & \multicolumn{2}{c|}{Dynamic Programming} & Automata & FM-Index \\ \cline{2-7}
                        & RADAR~\cite{Huangfu:DAC2018} & BioCAM~\cite{Karen:CORR2017}  & Race~\cite{Madhavan:ISCA2014} & RCAM~\cite{Kaplan:MICRO2017} & GenAx~\cite{Fuijiki:ISCA2018} & \textbf{FindeR} \\ \hline\hline
Die size ($mm^{2}$)     & 120  & 9.8K   & 450  & 383   & 4.6K  & 1.1K \\ \hline
Off-chip Memory(GB)     & 0    & 0      & 8    & 0     & 120   & 0    \\ \hline
Power(W)                & 12.5 & 153    & 24.3 & 6.6K  & 20    & 9.09 \\ \hline
Function                & \multicolumn{2}{c|}{Seeding} & \multicolumn{3}{c|}{Seed Extension} & Both \\ \hline
Throughput              & 125  & 186.8K & 2.1M & 177K  & 973   & \textbf{3.86M} \\ \hline
Throughput/Watt         & 10   & 1.2K   & 86K  & 26    & 48.65 &\textbf{424.6K}\\ \hline
\end{tabular}
\caption{The comparison between various accelerators for short read alignment.}
\label{t:hard_perf_all}
\vspace{-0.2in}
\end{table*}

\subsection{FM-Index for Short Read Alignment}

\subsubsection{Performance}

To highlight the performance of FindR on short read alignment, we compared it against prior hardware accelerators~\cite{Huangfu:DAC2018,Karen:CORR2017,Madhavan:ISCA2014,Kaplan:MICRO2017,Fuijiki:ISCA2018} focusing on aligning short reads. These accelerators can boost the performance of either seeding~\cite{Huangfu:DAC2018,Karen:CORR2017} or seed extension~\cite{Madhavan:ISCA2014,Kaplan:MICRO2017,Fuijiki:ISCA2018} for short reads. In contrast, only our FindeR is able to process both short read seeding by FM-Index searches and seed extension by FM-Index-based $k$-mismatch searches. 

The 3D-ReRAM-based RADAR~\cite{Huangfu:DAC2018} can only process EPMs with no mismatches, while the 2D-ReRAM-based PIM BioCAM~\cite{Karen:CORR2017} can create seeds with $\leq$$1$ mismatch. Both of them rely on huge capacity ReRAM CAMs consuming significant power and area overhead. With the same capacity, compared to memory arrays, CAM increases the power and area by $>10\times$~\cite{Li:ICCAD2016}. FindeR improves the short read seeding throughput and throughput per Watt by $20.7\times$ and $353.8\times$ over BioCAM. The more mismatches FindeR has to process, the more iterations it has to run. Therefore, FindeR attains 1179.4K, 707.7K and 424.6K 100-bp reads/sec/Watt with 0, 1 and 2 mismatches. When considering only EPMs, FindeR improves the short read seeding throughput per Watt by 117K$\times$ over 3D ReRAM-based RADAR. The significant improvement comes from two factors: First, the FM-Index algorithm of FindeR is much more efficient than the na\"{\i}ve exhaustive search adopted by RADAR; Second, the ReRAM CAMs have to consume more power than our RHUs, due to their huge peripheral circuits.

Both dynamic-programming-based accelerators (i.e., RaceLogic~\cite{Madhavan:ISCA2014} and RCAM~\cite{Kaplan:MICRO2017}) and the automata-based GenAx~\cite{Fuijiki:ISCA2018} require huge capacity for off-chip or internal memories to store complex intermediate data structures such as seed pointer tables, a position table, pointers, reads and references. They obtain optimal alignment results and can tolerate any number of mismatches by executing the high computational complexity SW algorithm. But FindeR improves the short read seed extension throughput and throughput per Watt by $83.8\%$ and $4.9\times$ over RaceLogic, respectively.

\subsubsection{Seed Quality}

Among all prior accelerators, GenAx achieves the highest quality~\cite{Fuijiki:ISCA2018} of short read seeding, since it extends a $k$-bp hashed seed by a stride of $k/2^n$-bp until a SMEM is built, where $n=[1,\lfloor log_2(k)\rfloor]$. To evaluate the seed quality, we implemented the GenAx hash-table-based SMEM seeding to generate seeds, and used BWA-MEM to produce alignment mappings on our CPU baseline. Because GenAx uses $k=12$~\cite{Fuijiki:ISCA2018}, the SMEM length of GenAx is often $\leq24$. FindeR creates SMEMs with an average length of 48. The alignment mappings built with the SMEM seeds generated by GenAx achieve 99.45\% sensitivity and 99.71\% precision. FindeR improves the sensitivity by 0.54\% and the precision by 0.28\%. Compared to GenAx, FindeR also reduces the hit number per read by 81\% in short read alignment. A smaller hits/read means FindeR less frequently invokes computationally expensive FM-Index-based $k$-mismatch searches and reduces the seed extension time. 

\begin{table}[htbp!]
\setlength{\tabcolsep}{2pt}
\centering
\begin{tabular}{|l|c|c||c|c|c|}
\hline
          & \multicolumn{2}{c||}{Performance}             & \multicolumn{3}{c|}{Quality}    \\\cline{2-6}
          & Throughput           & Throughput/Watt        & hits/read    &Sensitivity    & Precision \\\hline\hline
Dar-Pac   & \multirow{2}{*}{3.9K}& \multirow{2}{*}{0.26K} & 0.33         & 99.71\%       & 99.91\%     \\\cline{1-1}\cline{4-6}
Dar-ONT   &                      &                        & 0.45         & 98.2\%        & 99.1\%      \\\hline
Fin-Pac   & S: 2.9K              & S: 1.64K         	    & 0.29         & 99.8\%        & 99.95\%     \\\cline{1-1}\cline{4-6}
Fin-ONT   & E: 0.87K	           & E: 0.49K 	            & 0.44         & 98.31\%       & 99.23\%     \\\hline
\end{tabular}
\caption{The throughput and quality of long read seeding.}
\label{t:compare_to_darwin}
\vspace{-0.3in}
\end{table}

\subsection{FM-Index for Long Read Alignment}

We present the seeding results of long read alignment by comparing FindeR to a recent long read genomics co-processor Darwin~\cite{Turakhia:ASPLOS2018} in Table~\ref{t:compare_to_darwin}. Darwin proposes a novel hash-table-based seeding filter to reduce the seed extension overhead and a fast dynamic programming algorithm for APM in seed extensions. We used FindeR for long read SMEM seeding. Due to the large power-hungry SRAM buffers, Darwin consumes $15W$ power. So FindeR improves the throughput per Watt (the value having prefix ``S:'') by $5.3\times$ when only running for long read SMEM seeding. We implemented and executed the Darwin alignment flows on our CPU baseline to evaluate the seed quality. Compared to BWA-MEM and SGA, Darwin improves the sensitivity and precision of alignment and assembly mappings by $1\%\sim5\%$~\cite{Turakhia:ASPLOS2018}. However, unlike short read alignment relying only on the seed-and-extend paradigm, long read alignment requires error corrections for high error rate long reads~\cite{Wang:BMC2018}. When the FM-Index-based error correction technique FMLRC \cite{Wang:BMC2018} is applied, FindeR achieves slightly better sensitivity and precision ($<0.1\%$) than Darwin for long read alignment. But when FindeR performs error correction and SMEM seeding concurrently for long reads, its throughput per Watt (the value having prefix ``E:'') improvement over Darwin degrades to 88\%, due to the conflicts between error correcting and seeding requests inside each FindeR bank. Thanks to error corrections and SMEMs, FindeR decreases hits/read by $3\%\sim10\%$ over Darwin.

\section{Prior art}
\label{s:related_work}

Genome sequencing is the key to improving healthcare diagnoses, ensuring global food security and enforcing the wildlife conservation. But it is challenging to quickly and power-efficiently process such huge amounts of genomic data generated by genome sequencing machines. Application-specific acceleration for genome sequencing has become essential. 

Both approximate pattern matching (APM) and exact pattern matching (EPM) are essential to genome sequencing, due to the seed-and-extension computing paradigm. Because of the high time complexity of APM algorithms for seed extensions, recent work presents ASIC-~\cite{Madhavan:ISCA2014,Turakhia:ASPLOS2018,Fuijiki:ISCA2018}, FPGA-~\cite{Enzo:ICBB2017} and PIM-based~\cite{Huangfu:DAC2018,Kaplan:MICRO2017,Karen:CORR2017} accelerators to enhance the APM performance during seed extensions. Particularly, several works~\cite{Huangfu:DAC2018,Kaplan:MICRO2017,Karen:CORR2017} take advantage of power-hungry and low-density ReRAM-based content address memories (CAMs) to implement the SW algorithm to align short reads. However, compared to commodity ReRAM arrays, ReRAM CAMs with the same capacity enlarge the chip area and power consumption almost by $10\times$~\cite{Li:ICCAD2016}.

In this paper, we show that compared to APMs, EPMs are more elementary and more time-consuming in most critical steps of genome analysis. During the short read seed extensions, FM-Index-based $k$-mismatch searches can even completely replace APMs, because the majority of sequencing errors are substitutions~\cite{Quail:BMC2012} (mismatches). For EPM operations, prior work only creates FPGA-~\cite{Chang:FCCM2016,Houtgast:SAMOS2015,Arram:FPT2015,Arram:TCBB2017} and ASIC-based~\cite{Wu:ITBCS2017} accelerators to accelerate FM-Index searches notorious for massive random memory accesses. To the best of our knowledge, FindeR is the first PIM to accelerate FM-Index searches by ReRAM arrays.

\section{Conclusion}
\label{s:con}

Enhancing the computing efficiency of genome analysis is urgent and important for personalized medical care, since it is projected that each individual's genome may be sequenced and analyzed over the next decade. Unlike previous genomics accelerators, we focus on the hardware acceleration of EPM during genome analysis. We propose a reliable and power-efficient ReRAM-based Hamming distance unit to accelerate the FM-Index-based EPM widely adopted in critical steps of genome analysis including genome assembly, alignment, annotation and compression. We further architect a full-fledged pipelined PIM, FindeR, with a system library to improve FM-Index search throughput. Compared to prior accelerators, FindeR improves the FM-Index search throughput by $83\%\sim 30K\times$ and throughput per Watt by $3.5\times\sim 42.5K\times$.

\bibliographystyle{IEEEtran}
\bibliography{genomics}

\end{document}